# Deformation mechanisms of L-PBF-processed Ti-6Al-4V investigated using a combined experimental and simulation approach

Pushkar Prakash Dhekne [a], Nikhil Prabhu [a, b], Matthias Bönisch [a], Marc Seefeldt [a], Martin Diehl [a, b], Kim Vanmeensel [a]

[a] Department of Materials Engineering, KU Leuven, Kasteelpark Arenberg 44, Leuven B-3001, Belgium

[b] Department of Computer Science, KU Leuven, Celestijnenlaan 200A, 3001, Leuven, Belgium

## Abstract

Despite the significant application potential of laser powder bed fusion (L-PBF) processed Ti-6Al-4V components, a detailed understanding of their deformation mechanisms remains limited. This study investigates the deformation behavior of the α′ and α phases in the as-built and heat-treated specimens, respectively, using in-situ high-energy X-ray diffraction (HEXRD) combined with crystal plasticity modeling. Both phases exhibited similar elastic anisotropy, with the highest modulus along {00.2} and the lowest along {10.0}, although the α phase consistently showed higher directional moduli than the α′ phase. Their plastic deformation responses differed markedly: in the as-built α′ phase, slip activation followed the sequence prismatic → basal → pyramidal I <c+a>, whereas in the heat-treated α phase, the sequence was basal → prismatic → pyramidal I <c+a>. Analyses of full width at half maximum (FWHM) and diffraction peak intensities further supported these observations. Finally, inverse modeling within a crystal plasticity framework was employed to determine slip family–specific critical resolved shear stresses (CRSS), revealing higher CRSS values in the α′ phase for all slip systems except the prismatic family.

## 1. Introduction

Since the past decade, there has been a growing demand for laser-powder bed fusion (L-PBF) processed Ti-6Al-4V components from the automobile, aerospace, and biomedical industries [1–3]. This surge is attributable to the various advantages that L-PBF processing offers, such as on-site and on-demand manufacturing capability, reduced tooling cost, relative ease of manufacturing complex-shaped parts, and recyclability of the powder feedstock [4–6]. However, the high cooling rates generated during L-PBF processing trigger the martensite (α′) transformation and induce tensile residual stresses in Ti-6Al-4V [5]. Moreover, the spatially varying thermal history during L-PBF processing induces microstructural heterogeneity in the built component, especially in complex-shaped components [7–9]. Thus, a post-heat treatment is always needed to address these issues and optimize the strength-ductility combination. To date, various post-heat treatments have been proposed in the literature to optimize the strength-ductility balance of as-fabricated components [10–12]. Among these numerous heat treatments, those resulting in a bimodal microstructure have shown a better combination of quasi-static mechanical properties [13]. Recently, a two-step heat treatment was proposed to induce a bimodal microstructure, thereby improving the strength-ductility combination during quasi-static mechanical loading [14,15]. However, limited know-how about the deformation behavior of L-PBF-processed Ti-6Al-4V has limited its widespread adoption across various industries.

Various aspects of the deformation behavior of L-PBF-processed Ti-6Al-4V remain unknown, such as the ease of activation of different slip systems, or in other words, the critical resolved shear stress (CRSS) of the slip systems, and load partitioning between different grain orientations. Various datasets for critically resolved shear stress in Ti-6Al-4V have been reported in the literature, typically determined through various methods, such as micro-pillar compression tests, tensile deformation of single-crystalline samples, inverse modelling of macroscopic deformation response, and slip trace analysis during uniaxial tensile tests [16–20]. However, each of these methods has inherent limitations. CRSS values derived from micro-pillar compression are influenced by the dimensions of the pillar, and these tests are usually performed on pillars extracted from single grains, neglecting the effect of neighboring grains and thus failing to represent the deformation behavior of a polycrystalline microstructure [21]. In the case of slip trace analysis, the manual identification of slip traces limits the number of grains that can be examined. Furthermore, not all active slip systems produce visible surface traces, and diffuse markings can make identification more difficult [22]. The presence of a free surface may also affect deformation behavior [23–25]. Lastly, estimating CRSS values through inverse modeling of a single macroscopic tensile curve often results in non-unique solutions, since many parameters are fitted to a limited dataset.

To overcome many of these limitations, high energy X-ray diffraction (HEXRD) offers a powerful alternative. With its high flux, deep penetration, and fast detector sampling frequency, HEXRD enables the real-time capture of dynamic material processes such as the activation of deformation mechanisms, phase transformations, and chemical reactions [26–29]. In-situ HEXRD has been successfully integrated with both static and dynamic mechanical testing to investigate key aspects of deformation behavior, including the identification of active slip systems, load partitioning between phases, texture evolution, and the quantification of dislocation densities in metallic alloys [30–34]. However, HEXRD has seldom been applied to study deformation mechanisms in conventionally produced Ti-6Al-4V, let alone in its L-PBF-processed counterpart. Moreover, the orientation-specific deformation response extracted from diffraction patterns, when combined with crystal plasticity simulations, can provide insights that are otherwise inaccessible through diffraction data alone, such as the critical resolved shear stresses (CRSS) for different slip families or the identification of active slip systems in differently oriented grains [35].

Thus, the present study presents an integrated experimental-simulation approach to (A) investigate the elastic and plastic anisotropy for L-PBF-fabricated Ti-6Al-4V in both as-built and heat-treated states, (B) determine the CRSS for different slip families in α and α′ phases, and (C) explain the origin of the different deformation mechanisms in the two hexagonal phases, i.e., the hexagonal α' and α phases in as-fabricated and heat-treated specimens, respectively. The aims will be achieved by performing a uniaxial tensile test with in-situ HEXRD to investigate elastic and plastic anisotropy in both samples. Next, inverse modeling will be adopted to calibrate material parameters to determine the elastoplastic behavior of a representative volume element within a full-field crystal plasticity model. Here, a phenomenological model based on a power law relationship between the applied stress and the resulting plastic strain response will be used.

## 2. Materials and Methods

### 2.1. L-PBF processing

The Ti-6Al-4V tensile specimens were built on a 3D System ProX320 machine from Ti-6Al-4V (Grade 23) powder supplied by Carpenter Additive. The L-PBF parameters used are presented in Table 1. The tensile samples were fabricated horizontally with the tensile loading direction perpendicular to the build direction.

Table 1: L-PBF parameters applied to manufacture Ti-6Al-4V tensile bars

| Parameter | Value |
|---|---|
| Laser power (W) | 120 |
| Laser scanning speed (mm/s) | 1000 |
| Hatch spacing (μm) | 70 |
| Layer thickness (μm) | 30 |

## 2.2. Heat Treatment

The two-step heat treatment was performed in a vertical tube furnace under a protective Ar atmosphere. The first step involved heating to 940°C at 10°C/min, and the dwell time was set to 4 hours, followed by water quenching. The second step involved heating to 800°C at 10°C/min, and the dwell time was 2 hours, followed by slow furnace cooling (<3°C/min). A bimodal microstructure produced through such a two-step heat treatment has been shown to significantly enhance the mechanical properties of as-fabricated Ti-6Al-4V components [15].

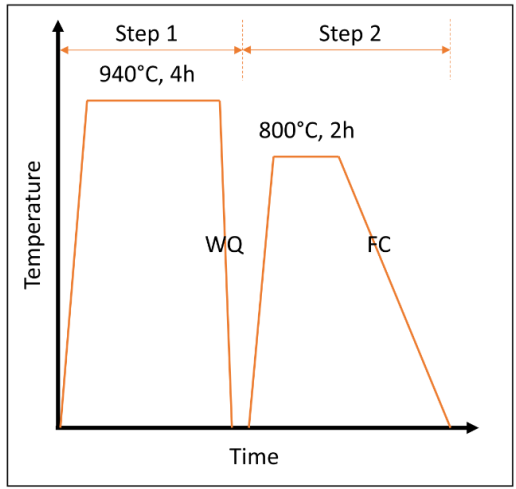

Figure 1: Two-step heat treatment to achieve the bimodal microstructure

## 2.3. In-situ high energy X-ray diffraction (HEXRD) experiment

The in-situ HEXRD experiment was performed at DESY on the PETRA III P07 beamline using a synchrotron X-ray beam of 100 keV. A schematic of the experimental setup is shown in Figure 2 (A). The tensile samples were loaded with a 10 kN load cell at a strain rate of 0.0003 s$^{-1}$, and the macrostrain strains were measured using a Bähr/TA Instruments DIL805A/D dilatometer. 2D diffraction patterns were continuously recorded on a Perkin Elmer detector (2048 x 2048 pixels) during the tensile test with an exposure time of 1s. The diffraction patterns for the as-built and heat-treated samples in the undeformed state are presented in Figure 2 (B & C), respectively. The detector tilt, sample-to-detector distance, beam center, and instrumental broadening were determined based on the diffraction pattern of a LaB$_6$ powder standard.

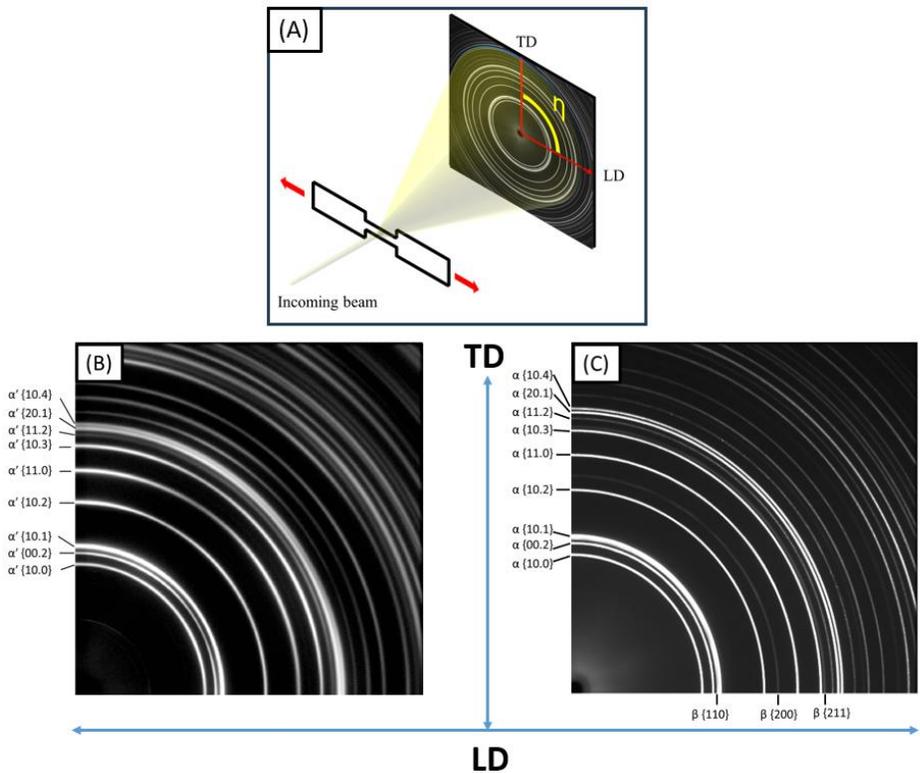

Figure 2: (A) Schematic of the in-situ tensile loading setup used at Petra III P07 of DESY. The 2D diffraction patterns for (B) as-built and (C) heat-treated specimens. Moreover, the position of the first nine reflections of the α and α' phases, as well as the three reflections of the β phase, are indicated.

## 2.4. Analysis of diffraction patterns

Each diffraction pattern was initially sectioned into 72 wedges with a 5° width. The lattice strains on different crystallographic planes were calculated by performing Rietveld refinement of the entire diffraction pattern in MAUD [36]. The Delft type line broadening and isotropic size strain model were used during the Rietveld refinement. The extended Williams-Imhof-Matthies-Vinel (E-WIMV) texture model was used to simulate crystallographic texture [37]. The radial diffraction in the diamond anvil compression (rDAC) strain model was employed to model the lattice strains [38,39]. The interplanar distance in the loading and transverse directions for various crystallographic planes was calculated by Eqn. (1) and (2)

$$d_{\parallel\{hk.l\}} = d_p[1 + (1 - 3\cos^2\psi)Q_{\{hk.l\}}] \quad \text{Eqn. 1}$$

$$d_{\perp\{hk.l\}} = d_p(1 + Q_{\{hk.l\}}) \quad \text{Eqn. 2}$$

Where $d_{\parallel(hk.l)}$ represents the average interplanar spacing of {hk.l} planes in grains with their {hk.l} plane normal parallel to the loading direction. Similarly, $d_{\perp(hkl)}$ indicates the average interplanar spacing of {hk.l} planes in the grains with their {hk.l} plane normal parallel to the transverse direction. $\psi$ is the angle between the loading axis and the diffraction plane normal. $Q_{\{hk.l\}}$ is the fitting parameter of rDAC model. The lattice plane-specific strain in both loading and transverse directions is calculated from Eqn. 3

$$\epsilon_{hk.l} = \frac{d - d_0}{d_0} \quad \text{Eqn. 3}$$

Here, $d_0$ is obtained by refining the diffraction pattern obtained before deformation.

## 2.5. Inverse Modeling

The elastoplastic mechanical responses exhibited by the two samples were simulated with the help of full-field simulations. An agreement between experimental and simulated responses was sought through iteratively calibrating material parameters of the underlying constitutive equations. To this end, an optimization algorithm was adopted. The following sub-sections provide the details of the inverse modelling approach.

### 2.5.1. Crystal Plasticity Simulations

A spectral solver within DAMASK, an open-source multi-physics material behavior simulation toolkit, was employed to simulate the elastoplastic response of the samples [40]. DAMASK is formulated in a large-strain framework, in which the total deformation gradient tensor $F$ of each individual material point is multiplicatively decomposed into a plastic isochoric deformation gradient tensor $\mathbf{F_p}$ that maps from reference configuration to plastic configuration and elastic non-isochoric deformation gradient tensor $\mathbf{F_e}$ which maps from the intermediate to the deformed configuration. The elastic Green-Lagrange strain tensor $\mathbf{E_e} := (\mathbf{F_e}^T \mathbf{F_e} - \mathbf{I})/2$ and its work-conjugated second Piola–Kirchhoff stress tensor $\mathbf{S}$ are related via Hooke's law

$$\mathbf{S} = \mathbf{C}\mathbf{E_e} \qquad \text{Eqn. 4}$$

where $\mathbf{C}$ is the fourth-order stiffness tensor.

The evolution of the plastic deformation gradient follows $\mathbf{\dot{F}_p} = \mathbf{L_p}\mathbf{F_p}$, where $\mathbf{L_p}$ is the plastic velocity gradient. In crystal plasticity, $\mathbf{L_p}$ comprises the contributions exclusively from slip systems and is defined as a projected sum of shear rates $\dot{\gamma}$ on these slip systems

$$\mathbf{L_p} = \sum_{i=1}^{N_s} \dot{\gamma}^i \mathbf{P}^i \qquad \text{Eqn. 5}$$

where $N_s$ is the total number of slip systems and $\mathbf{P} := \mathbf{s} \otimes \mathbf{n}$ is the Schmid matrix of a slip system constructed from its plane normal n and slip direction s.

A constitutive model at the slip-system level defines the relationship between the applied stress $\mathbf{M_p}$ (Mandel stress in plastic configuration) and the plastic response $\dot{\gamma}$. When non-Schmid effects are discarded, the applied stress can be resolved onto a slip system as

$$\tau = \mathbf{M_p} \cdot \mathbf{P} \qquad \text{Eqn. 6}$$

where $\tau$ is the resolved shear stress that drives plastic deformation in the material. A widely used phenomenological model, where the material state is characterized by resistance offered to plastic slip, has been adopted [41]. A slip system-specific slip resistance, $\xi$, evolves from its initial value $\xi^0$ to a saturation value $\xi^\infty$ according to

$$\dot{\xi}^i = h_0{}^i \left|1 - \frac{\xi^i}{\xi_\infty{}^i}\right|^a sgn\left(1 - \frac{\xi^i}{\xi_\infty{}^i}\right) \sum_{j=1}^{Ns} h^{ij}|\dot{\gamma}^j| \qquad \text{Eqn. 7}$$

where $h_0$ and $a$ are slip family-dependent fitting parameters, governing the initial slope and shape, respectively, of the hardening curve. Eqn. (7) shows that apart from self-hardening, the hardening behavior of a slip system $\alpha$ is also dependent on the neighboring slip systems $\alpha'$ and these inter- and intra-slip family interactions is defined by elements of the interaction matrix $h^{\alpha\alpha'}$. The plastic response, $\dot{\gamma}$, of a slip system $\alpha$, is then dependent on its current resistance to slip $\xi^\alpha$ and the resolved shear stress $\tau^\alpha$ through a power law-like relation

$$\dot{\gamma}^i = \dot{\gamma}_0 \left|\frac{\tau^i}{\xi^i}\right|^n sgn(\tau^i) \qquad \text{Eqn. 8}$$

where $\dot{\gamma}_0$ and $n$ are the reference shear rate and the inverse of the rate sensitivity exponent, respectively.

### 2.5.2. Computation of Lattice Strains

Before extracting the elastic strain for the different lattice planes, grains that satisfied Bragg's diffraction condition for each lattice plane were identified. The grains whose lattice plane normal, including symmetrically equivalent ones, aligned within a threshold angle of 3.5 degrees with the loading direction were considered to satisfy the diffraction condition. Once the grains pertaining to a plane family were identified, the elastic component of the total strain in these grains along the plane normal was computed. These elastic strains were then averaged to yield the lattice strain of the corresponding plane family.

### 2.5.3. Material Parameters Calibration

The elastic and the plastic parameters for the constitutive model were calibrated separately against the experimental data set. Firstly, the single crystal elastic constants were calibrated against the elastic regime of the lattice strain vs global stress data for the four lattice planes ({10.0}, {00.2}, {10.1}, and {10.2}), corresponding to the first four reflections of the hexagonal phases. Secondly, the plastic parameters were calibrated by fitting a part of the experimental dataset, starting from just below the yield stress till the ultimate tensile strength. Here, the plastic response of the material has been represented by considering only three slip systems, viz. basal <a>, prismatic <a>, and pyramidal I <c+a>. However, the interactivity of the three slip families renders difficulties in calibrating model parameters, as even a relatively simple phenomenological model ends up having a huge number of parameters with interdependencies.

To tackle this challenge, the calibration was carried out systematically not just against a single global stress-strain response, but also against the orientation-specific mechanical response, capturing the disparity in slip family contribution to the overall behavior. Hence, the plastic parameters in the present approach were calibrated in two steps: in the first step, only the global stress-global strain dataset was considered as the reference. In the second step, in addition to the global dataset, global stress-lattice strain datasets of the four plane families were also considered as reference. A direct search Nelder-Mead optimization algorithm was adopted to find optimal values of model parameters.

The routines for material parameter identification were implemented in Python and utilized a variant of the Nelder-Mead simplex algorithm [42] with adaptive parameters, incorporated in a Python library for scientific computing SciPy [43]. The Nelder-Mead simplex algorithm is a local non-gradient-based optimizer whose ability to converge at an optimal point is sensitive to the initial guess values of the parameters.

### 2.5.4. Formulation of Loss Functions

The optimizer, for each iteration, computes losses by juxtaposing the reference experimental data (exp) against the simulated material responses (sim). As mentioned in the previous section, the five independent elastic constants of hexagonal symmetry of the linear elastic model were calibrated against the experimental global stress vs lattice strains data set. The discrepancy from the reference was quantified as a scalar loss value according to:

$$L_\text{e} = \sum_{\{hk.l\}} \frac{\left\| \epsilon_{\text{LS,exp}} - \epsilon_{\text{LS,sim}} \right\|_2}{N} \qquad \text{Eqn. 9}$$

where N is the total number of data points, and LS stands for lattice strains. $\epsilon$ represents the lattice strain of a plane family, and $\|.\|2$ is the $l^2$-norm.

In the first "global" step of the plastic parameter calibration routine, the total loss $L_{p,G}$ -where the subscripts p and G stand for plastic and global, respectively- comprises the contributions from the actual values (0) and their first derivatives (1) and reads

$$L_\text{p,G} = w_\text{p,G0} L_\text{p,G0} + f_\text{G} w_\text{p,G1} L_\text{p,G1} \qquad \text{Eqn.10}$$

The two contributions were weighted relatively ($w_{p,G0}$ and $w_{p,G1}$) and a factor $f_G$ was calculated during the first iteration to consistently bridge the difference in magnitude between $L_{p,G0}$ and $L_{p,G1}$.

In the second step, in addition to the global response of the sample, lattice strains of the four plane families together constitute the reference data set. For this case, the total plastic loss function consisted of two additional terms-on the top of Eqn (10) – $L_{p,LS0}$ and $L_{p,LS1}$, and two more factors to bridge differences in magnitudes. The total loss of plastic fitting $L_p$ then reads

$$L_p = w_{p,G} L_{p,G} + f_{LS} w_{p,LS} L_{p,LS} \qquad \text{Eqn. 11}$$

$L_{p,LS}$ is computed in a similar way as displayed in Eqn. (9), however, now also including the first derivative.

### 2.5.5. Initial Guess Values and Constraints

The initial guess values were chosen carefully as the initial simplex significantly affects the search direction within the Nelder-Mead method. For the elastic parameters, initial values were taken from ref. [42]; a minor difference in the present approach is that $C_{33}$ was taken as an independent parameter.

In the first step of the plastic parameters' optimization, only three parameters were considered independent. These were $a_p$, $h_0$ and $\xi_{0,p}$ representing model-specific fitting exponent, initial hardening rate and initial slip resistance, respectively, of the prismatic slip family. The parameters pertaining to the basal and pyramidal slip families were deemed to be dependent on the prismatic slip family parameters and were kept constant by keeping the ratios $\frac{\xi_{0,b}}{\xi_{0,p}}$ and $\frac{\xi_{0,\pi}}{\xi_{0,p}}$ fixed. Here, a value of 1.05 was chosen for $\frac{\xi_{0,b}}{\xi_{0,p}}$ [43]. While two different values were chosen for the $\frac{\xi_{0,\pi}}{\xi_{0,p}}$, 1.68 for the heat-treated specimen and 3 for the as-built specimen [43,44]. The different values were chosen due to the significantly different behavior of {00.2} grains in both specimens and these grains tend to deform via the pyramidal I ⟨c+a⟩ slip system. In the as-built sample, higher load bearing capacity of {00.2} grains motivated to choose a higher value for $\xi_{0,\pi}$ compared to the heat-treated specimen. The initial value for $\xi_{0,p}$ was chosen to be $\frac{\sigma_y}{2.65}$, where 2.65 is the average Taylor factor from iso-stress and iso-strain conditions. Furthermore, a multiplier $\frac{\sigma_{max}}{\sigma_y}$ was used on $\xi_0$ to obtain $\xi_\infty$. In the second and the final optimization step where additionally the lattice strain data was also included, the reference data now embodied the information about the disparity in plastic activity of different slip families. The fixed-ratio constraints that were imposed on the plastic parameters in the previous step were lifted, resulting in twelve independent parameters that are to be calibrated: a, $h_0$, $\xi_0$ and $\xi_\infty$ of each slip family.

## 3. Results & Discussion

### 3.1. Microstructure Analysis

Figure 3 (A, B) presents the microstructure of L-PBF processed Ti-6Al-4V in the as-built and heat-treated condition. The microstructure in the as-built state comprised the α′ phase, while the heat-treated microstructure was composed of 94% α and 6% β.

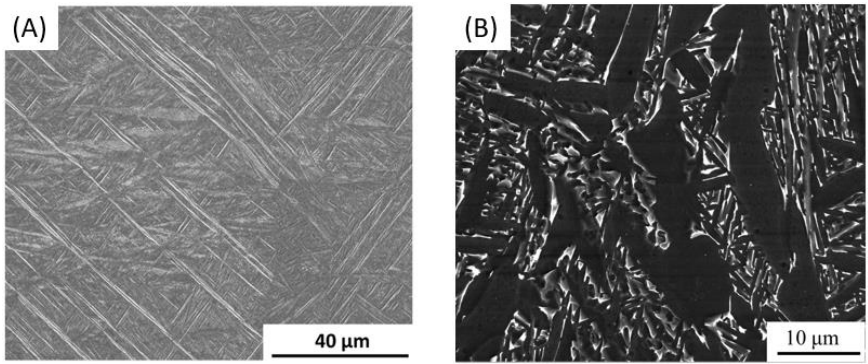

Figure 3: The microstructure of Ti-6Al-4V specimen in the (A) as-built and (B) the heat-treated conditions.

### 3.2. Macroscopic tensile properties

Figure 4 illustrates the macroscopic tensile curve for the as-built and heat-treated specimen obtained during the in-situ tensile test at DESY. As expected, the as-fabricated specimen exhibited higher ultimate tensile strength with reduced ductility compared to the heat-treated specimen [30]. However, a similar yield strength ($\sigma_{y,0.2}$) was observed for both samples.

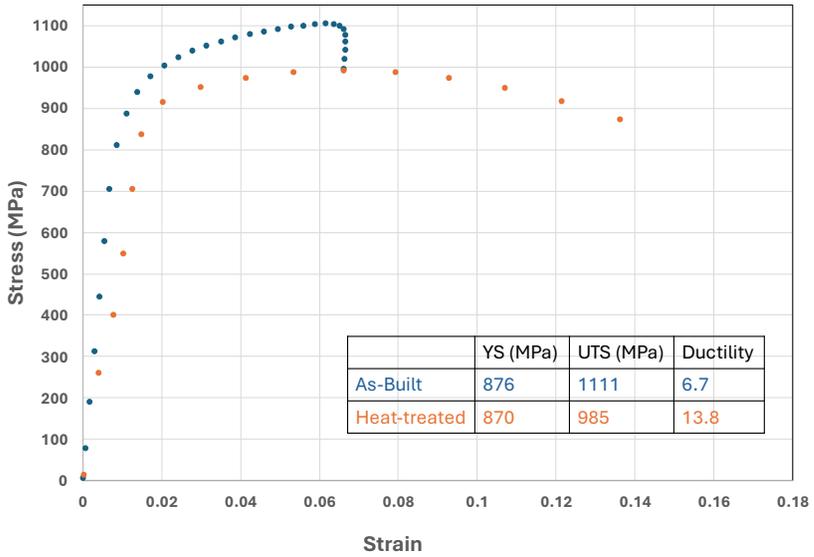

*Figure 4: Macroscopic true stress-strain curves for the as-built and heat-treated specimens obtained during in-situ tensile deformation performed in combination with HEXRD*

## 3.3. In-situ deformation monitoring with HEXRD

Figure 5 (A, B) displays the lattice strain evolution with the true stress for nine different grain families of the hexagonal α′ and α phases. For brevity, the lattice strain evolution for the β phase is not presented here but is discussed in section 3.3.2. Here, the positive and negative strains represent the strains in the loading and transverse directions, respectively.

In the α′ phase of the as-fabricated sample, the lattice strain along the loading direction in all grain families evolves linearly up to 800 MPa (a stress value below the macroscopic $\sigma_{y,0.2}$ yield point); thereafter, all curves deviate from the initial linear behavior. In contrast, the lattice strain of the α phase in the heat-treated sample evolves linearly up to its macroscopic yield point and the deviation from linearity is observed only after the yield stress is exceeded. The different behavior of the hexagonal α' and α phases in the as-fabricated and heat-treated samples, respectively, will be extensively discussed in later sections.

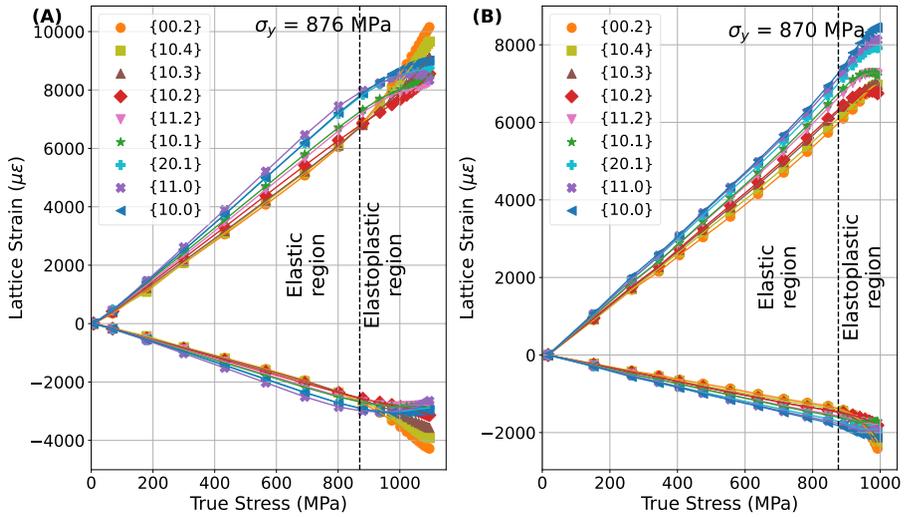

*Figure 5: Lattice strain evolution in (A) α′ phase of as-built and (B) α phase of heat-treated specimens. Positive strain values indicate tensile strains along the loading direction, and negative strain values denote compressive strains along the transverse direction. The dotted vertical line indicates the macroscopic yield point for the respective sample.*

### 3.3.1. Micro-mechanical Deformation Behavior of α′ phase in the As-Built State

*3.3.1.1. Elastic deformation*

Figure 6 illustrates the lattice strain evolution within the elastic regime for the α′ phase. The significant differences in the lattice strains across differently oriented grains reveal the elastic anisotropy of the α′ phase. Based on the lattice strain distribution, the {10.0} grains were observed to be the most compliant, while the {00.2} grains were the stiffest, revealing an inverse relationship between the apparent Young's modulus and the angle between the loading direction and the c-axis, as is commonly observed for hexagonal crystal structures [47].

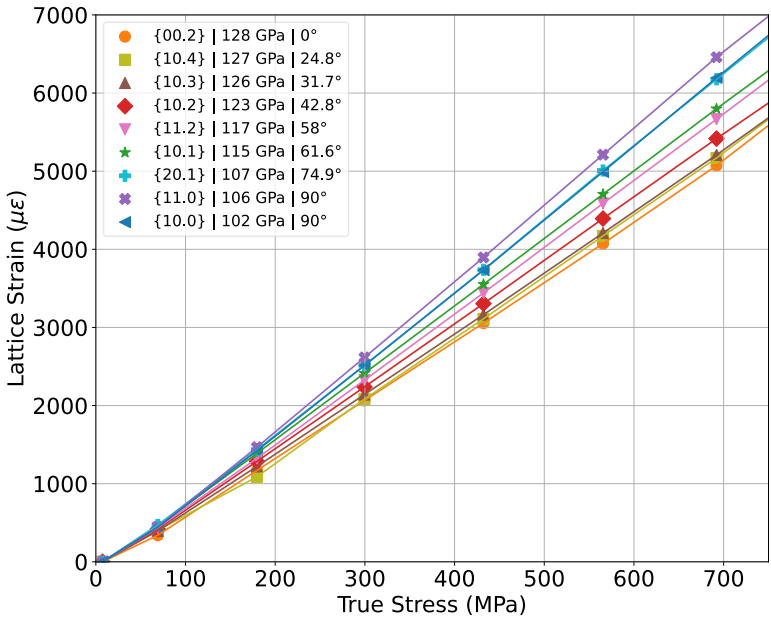

*Figure 6: Lattice strain evolution for nine reflections of the α' phase. The legend indicates the Youngs modulus (in GPa) and the inclination of the diffracting plane with the basal plane.*

*3.3.1.2. Elastoplastic deformation*

Figure 7 (A) shows that the initial deviation from linear behavior and reduction in slope occurs for {11.0} grains once the applied stress exceeds 850 MPa. This reduction in slope, or flattening of the lattice strain curve, signals the onset of plastic deformation in a grain family [48]. The initiation of plastic deformation in {11.0} grains indirectly suggest the activation of prismatic slip, as these grains have a higher Schmid factor for prismatic slip systems (see Appendix A) [49–51]. At the same time, an upward inflection in the lattice strain curves for {00.2}, {10.4}, and {10.3} grains is observed in Figure 7 (C), suggesting a load transfer from the {11.0} grains onto these grains. At the same time, the constant slopes of the lattice strain curve for {10.0}, {20.1}, {10.1}, {10.2}, and {11.2} grains indicate the continuation of elastic deformation in these grains (Figure 7 (A, B)). With further loading up to 910 MPa, plastic deformation initiates in the {10.0}, {20.1}, and {10.1} grains. Based on the Schmid factors presented in Appendix A, it can be predicted that the {10.0} and {20.1} grains predominantly deform via prismatic slip, while the plastic deformation in the {10.1} grains is accommodated by a combination of basal and prismatic slip.

Beyond 920 MPa, the {10.2} and {11.2} grains also deform plastically, primarily through the basal slip activation, based on Schmid factor analysis. Meanwhile, lattice strains in the {00.2}, {10.4}, and {10.3} grains continue to increase at an accelerating rate, indicating ongoing load transfer, without any flattening of the lattice strain curves, even up to the ultimate tensile strength (UTS). The absence of flattening in their lattice strain curves for {00.2} and {10.4} grains suggest a limited activation of the pyramidal I <c+a> slip system [52]. This peculiar lattice strain evolution will be further analyzed in section 3.4.3.

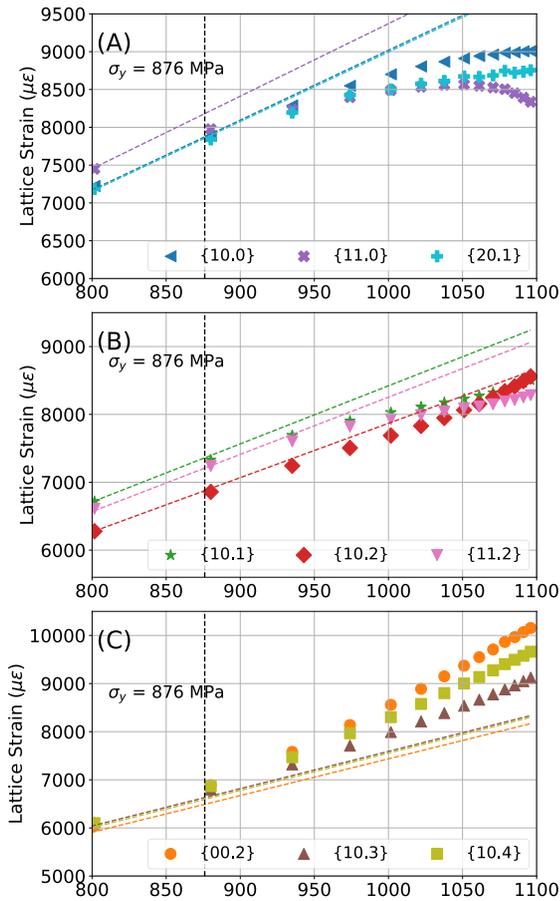

Figure 7: Lattice strain vs. true stress curves for the nine different grain families of the α' phase. The lattice strain curves for (A) {10.0}, {11.0}, {20.1}; (B) {10.1}, {10.2}, {11.2}; and (C) {00.2}, {10.3}, {10.4} are shown. In panels (A), (B), and (C), dotted lines of the same color as the data points represent linear fits to each grain family in the elastic region, highlighting the

*change in slope during the transition from elastic to plastic deformation. The vertical dotted line indicates the macroscopic yield stress.*

### 3.3.1.3. Full Width Half Maximum (FWHM) and Intensity Evolution

The evolution of the FWHM and intensity for the different α′ phase reflections is plotted in Figure 8 (A-F). Both parameters have been normalized with their respective values in the unloaded state. Generally, an increase in the FWHM of a diffraction peak may result from either a higher dislocation density or a heterogeneous strain distribution [32,53]. However, the impact of heterogeneous strain distribution on peak broadening is significantly less than that of increased defect density [54]. Therefore, a drastic increase in FWHM can serve as an indicator of plastic deformation [31]. On the other hand, intensity variation can give information about the grain rotation during plastic deformation, which can also indirectly indicate the active slip system.

- *Prismatic planes ({10.0}, and {11.0}): Figure 8(B, E)*

Figure 8 (B) shows that the {11.0} reflection broadens above 750 MPa. However, a drastic increase in FWHM is only observed after the stress level reaches 850-900 MPa, higher than the yield point of the {11.0} grains, as indicated by their lattice strain curves (Figure 7 (A)). In contrast to the {11.0} reflection, the {10.0} reflection behaves differently: it initially narrows by nearly 7% before broadening above 900 MPa, close to the yield point of the {10.0} grains. This initial narrowing in {10.0} grains could have resulted from dislocation annihilation. As both {10.0} and {11.0} reflections broaden above their respective yield points, the peak broadening can be attributed to increased dislocation density owing to the initiation of plastic deformation in these grains. Unlike the FWHM evolution, the {10.0} and {11.0} reflections display opposite trends in peak intensity: post-yield, the intensity of the {11.0} reflection decreases by almost 50%, whereas the intensity of the {10.0} reflection increases by 10%. Previous studies have linked this contrasting intensity trend for the {10.0} and {11.0} reflections to grain rotation due to prismatic slip activation in these grains [55].

- *Pyramidal planes ({10.1} and {10.2}): Figure 8 (A, D)*

{10.1} and {10.2} reflections slightly broaden above 750 MPa, which is significantly below their respective yield stresses as indicated by their lattice strain curves (Figure 7 (B)). After reaching their respective yield points, the FWHM of the {10.1} reflection increases drastically, while that of the {10.2} reflection decreases. The {10.1} and {10.2} reflections exhibit contrasting trends in intensity evolution: while the intensity of {10.1} increases, that of {10.2} decreases.

- *Basal or near-basal planes ({00.2}, {10.4}, and {10.3}): Figure 8 (C, F)*

The {00.2}, {10.4}, and {10.3} reflections exhibit a significantly different trend compared to the other reflections. All three reflections significantly broaden above 920 MPa, indicating the onset of plastic deformation, while their intensity decreases simultaneously. However, the respective lattice strain curves did not exhibit any downward inflection, indicating limited plastic deformation in these grains. This discrepancy needs further attention and will be addressed in section 3.4.3.

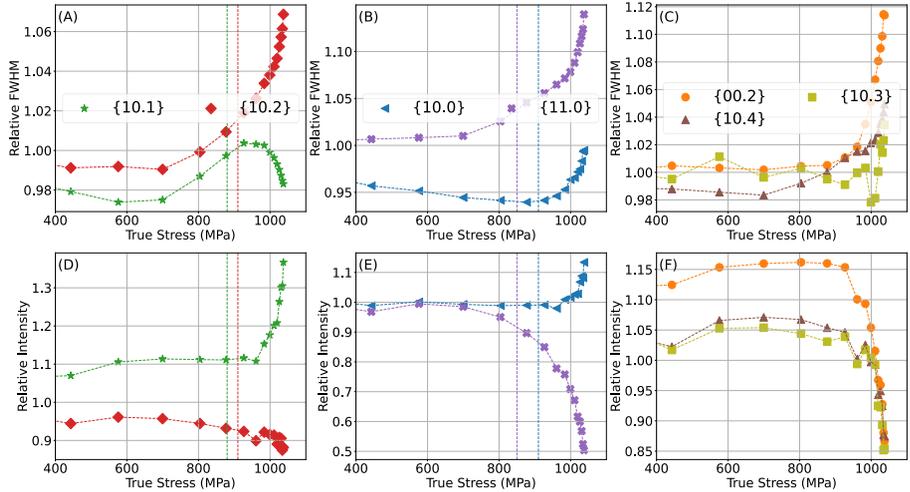

*Figure 8: Evolution of normalized FWHM and intensity is presented in (A, D) {10.1} and {10.2}; (B, E) {10.0} and {11.0}; (C, F) {00.2}, {10.3}, and {10.4}. The vertical dotted lines in (A-D) present the yield strength of different grain families. No dotted lines have been shown in (D & F) as no plastic yielding was observed for {00.2}, {10.3}, and {10.4} grains in the lattice strain curves.*

### 3.3.2. Micro-mechanical Deformation Behavior of α Phase in the Heat-treated Specimen

*3.3.2.1. Elastic Deformation*

Figure 9 presents the Young's modulus for the nine distinct grain families of the α phase in the heat-treated specimen. Similar to the α' phase in the as-built sample, the α phase in the heat-treated specimen demonstrates a similar inverse relationship between the Young's constant and the c-axis inclination of the loading direction. Notably, the α phase exhibits higher elastic constants than the α' phase, which could be attributed to its higher aluminum content [47].

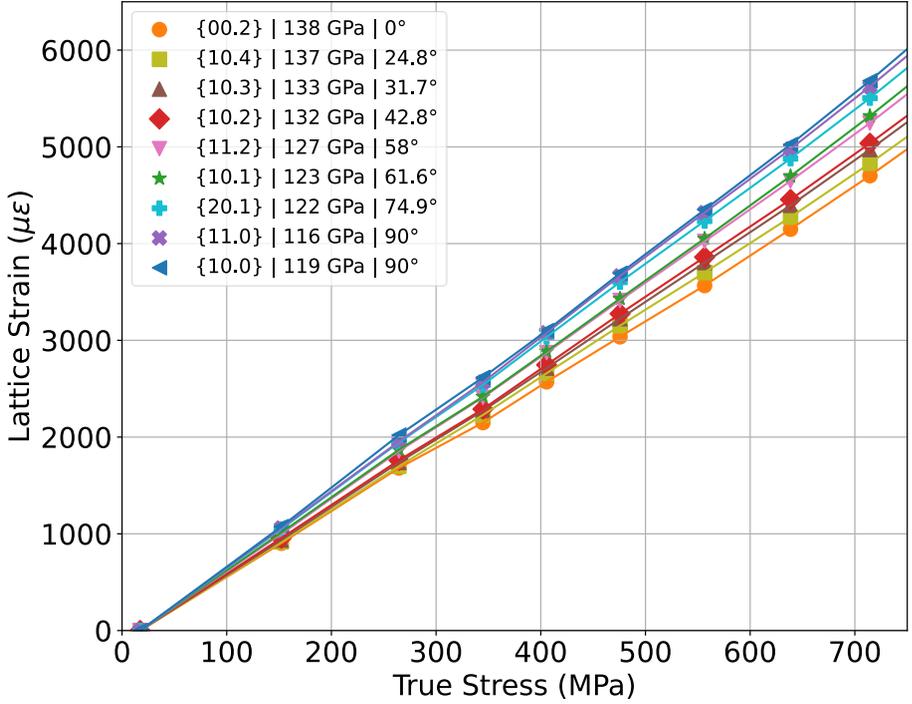

*Figure 9: Evolution of the lattice strain with the applied stress in the elastic region (0-750 MPa). The legend also indicates the Young's modulus and inclination angle of the diffracting plane with the basal plane.*

### 3.3.2.2. Elastoplastic deformation

A distinct plastic deformation mechanism was observed in the heat-treated state (Figure 10 (A–F)), compared with the as-built state. Plastic deformation of the heat-treated specimen initiates in the β phase, particularly within the {200} family of grains at 890 MPa. This deformation is facilitated by the activation of the {112}<111> slip system [56]. The plastic deformation in the {200}β grains induces load transfer to the {110}β, {10.0}α, {00.2}α, {11.0}α and, {20.1}α grain families, as evidenced by an upward inflection in their respective lattice strain curves. At 913 MPa, plastic deformation of the α phase commences in the {10.2} grains, primarily supported by the basal slip activation [57]. Upon further loading, plastic deformation initiates across different grain families such as {10.1}α, {10.0}α, {11.2}α, {10.3}α, {11.0}α, and {20.1}α at stress levels of 925, 945, 946, 947, 954, and 962 MPa, respectively. As seen earlier, the plastic deformation of the {10.0}α, {11.0}α, and {20.1}α grains suggests the activation of the prismatic slip. Whereas, the activation of basal slip is suggested by the plastic deformation of the {10.1} and {11.2} grains. Meanwhile, plastic deformation is also observed in the {110}β and {211}β grain families at 947 MPa. Finally, the lattice strain curves for the {00.2} and {10.4} grains exhibit a downward inflection at 970 MPa. Based on the Schmid factor analysis, the downward inflection in the lattice strain curves of the {10.4} and {00.2} grains indicates the activation of the pyramidal I <c+a> slip system.

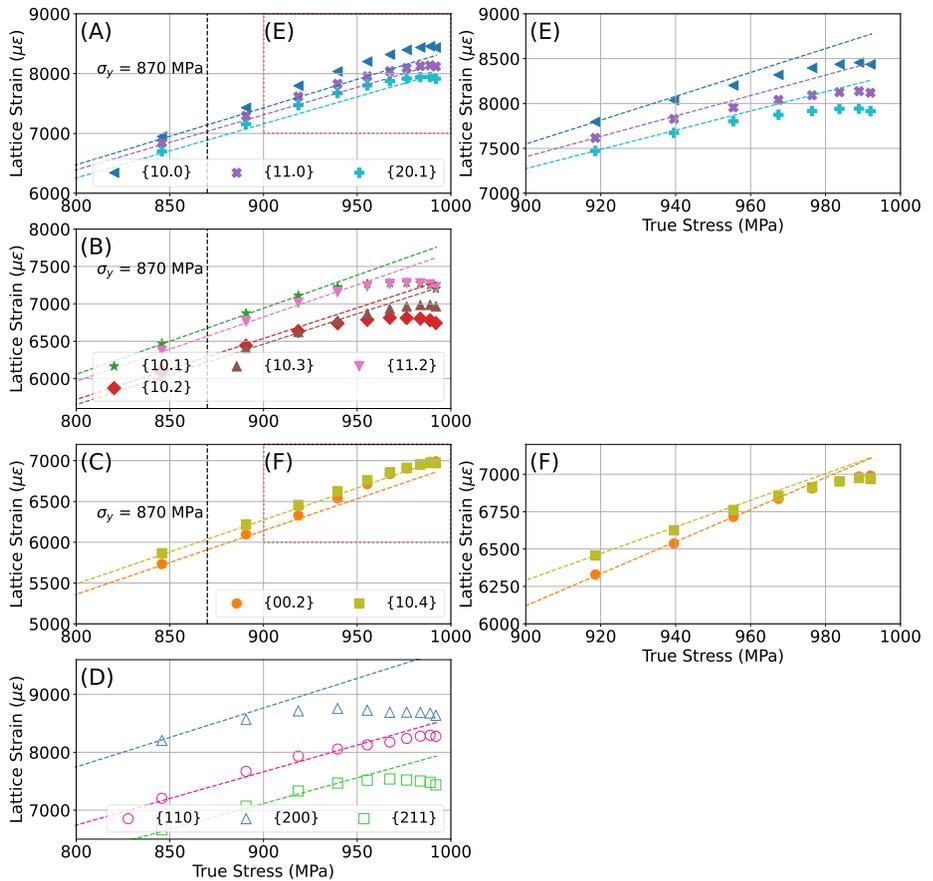

*Figure 10: Lattice strain evolution for α and β phases in the heat-treated specimen. The lattice strain evolution for (A) {10.0}α, {11.0}α, and {20.1}α; (B) {10.1}α, {10.2}α, {10.3}α, and {11.2}α; (C) {00.2}α and {10.4}α; (D) {110}β, {200}β, {211}β are presented. The dotted line represents the linear slope in the elastic region. The vertical lines in the panel (A) – (D) indicate the macroscopic yield point of the specimen. Panels (E) and (F) offer a zoomed-in view of the (A) and (C) to highlight the plastic deformation for {10.0}α, {11.0}α, and {20.1}α and {00.2}α and {10.4}α, respectively. The dotted line represents the linear regression of the first four data points after the upward inflection above 900 MPa. This is performed to indicate the plastic deformation of the aforementioned grains.*

3.3.2.3. Full Width Half Maximum (FWHM) and Intensity Evolution

Figure 11 depicts the FWHM and the intensity evolution for the different α phase reflections as a function of true stress. All diffraction peaks exhibit narrowing and intensification up to 850 MPa, a behavior that was not observed in the as-built state. Beyond 850 MPa, FWHM evolves differently across various α phase reflections. Although diffraction peak narrowing has been observed in the elastic regime, it has not been previously commented on in ref. [58]. The peak narrowing can be attributed to the reduced dislocation density in the elastic regime, caused by dislocation recombination, dislocation sinking into grain boundaries, and the formation of low-energy dislocation structures [32,59].

- *Pyramidal planes ({10.1}, {10.2}, {10.3}, and {11.2}): Figure 11 (A, D)*

In line with the lattice strain evolution, the {10.2} reflection is the first to broaden above 875 MPa, very close to its yield point, indicating that the increase in FWHM is attributable to the increased dislocation density. Further loading causes broadening of the {10.3} reflection at 910 MPa, and the {10.1} and {11.2} reflections at 925 MPa, close to their respective yield points. Conversely, the intensity of the {10.2} and {11.2} reflections decrease above their respective yield points. The intensity of the {10.1} diffraction signal initially decreases slightly but increases at higher levels of true stress.

- *Prismatic planes ({10.0} and {11.0}): Figure 11 (B, E)*

The {10.0} and {11.0} reflections of the α phase broaden considerably after their yield points, attributable to an increase in dislocation density in these grains. Post-yielding, the intensification of the {10.0} reflection and attenuation of the {11.0} reflection indicate the prismatic slip activation.

- *Basal or near-basal planes ({00.2} and {10.4}): Figure 11 (C, F)*

The {00.2} and {10.4} grains exhibit peak broadening at 870 MPa and 910 MPa, respectively, well below their yield stresses. Conversely, these peaks exhibit opposite intensity trends, with an increase in intensity for the {00.2} reflection and a decrease in intensity for the {10.4} reflection.

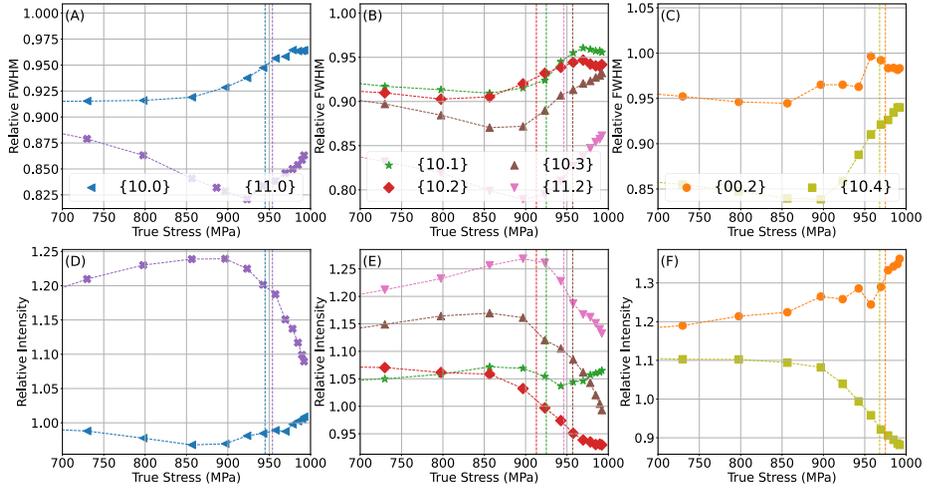

*Figure 11: Evolution of normalized FWHM and intensity is presented in (A, D) {10.1} and {10.2}; (B, E) {10.0} and {11.0}; (C, F) {00.2}, {10.3}, and {10.4}. The vertical dotted lines in (A-F) present the yield strength of different grain families.*

## 3.4. Crystal plasticity simulation

### 3.4.1. Elastic Region

Figure 12 (A, B) compares the experimental and simulated lattice strains in the elastic regime for the as-built and the heat-treated speciemn, showing that the modelling approach successfully captured the lattice strain evolution in the elastic region for both specimens. The optimized values of the five independent elements of the stiffness tensor for α′ and α phases are presented in Table 2; in the same table, the initial guess values of the optimization procedure are also reported. The optimized values are within the range of the reported elastic constants for the α phase in Ti-6Al-4V [44,46].

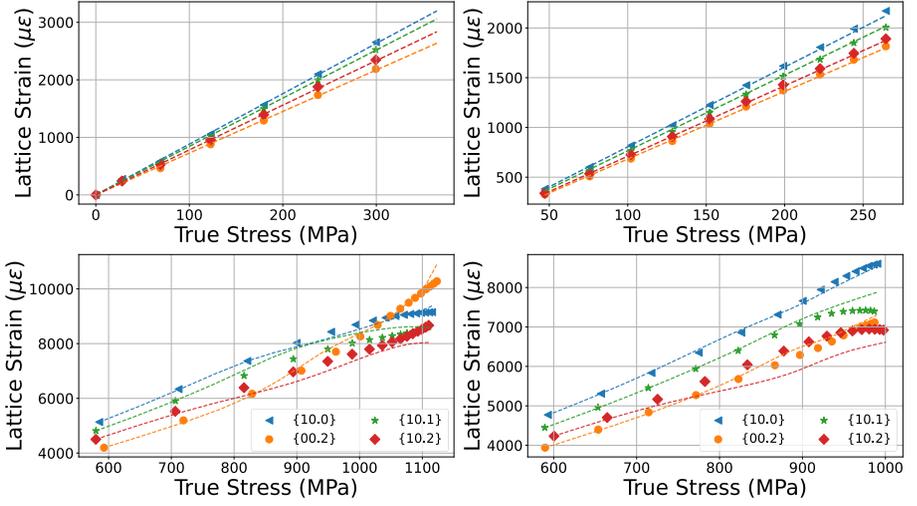

Figure 12: A comparison of the experimental and simulated lattice strains for the as-built specimen in (A) elastic and (C) plastic regions. Similarly, the comparison between the experimental and simulated lattice strains for the heat-treated specimen in (B) elastic and (D) plastic regions. The data points correspond to the experimentally determined lattice strain curves, while the dotted lines represent the simulated lattice strains calculated using the optimized parameters.

Table 2: Optimized elastic constants (in GPa) for the α' and α phases

|  | α' (As-Built) | α (Heat-treated) | Initial value |
|---|---|---|---|
| $C_{11}$ | 151 | 148 | 169 |
| $C_{12}$ | 83 | 61 | 89 |
| $C_{13}$ | 59 | 67 | 62 |
| $C_{33}$ | 187 | 203 | 196 |
| $C_{44}$ | 56 | 60 | 43 |

### 3.4.2. Plastic Region

Figure 12 (C, D) presents the experimental and simulated lattice strains in the plastic regime. In both cases, the model accurately captured the lattice strain evolution for {10.0} and {00.2} grains. However, an accurate quantitative

agreement between the experimental and simulated lattice strain evolution was not reached for {10.1} and {10.2} grains. The optimized set of plastic parameters is presented in Table 3. The α′ phase possesses higher CRSS for all slip systems, except the prismatic slip system, compared to the α phase.

*Table 3: Optimized plastic parameters*

| Slip Family | Parameter | α′ (As-Built) | α (Heat-treated) |
|---|---|---|---|
| Basal | $\xi_0$ (MPa) | 349 | 332 |
| | $\xi_\infty$ (MPa) | 409 | 334 |
| | $h_0$ (MPa) | 1428 | 2568 |
| | a | 0.36 | 0.99 |
| Prismatic | $\xi_0$ (MPa) | 312 | 347 |
| | $\xi_\infty$ (MPa) | 319 | 415 |
| | $h_0$ (MPa) | 5293 | 2905 |
| | a | 0.64 | 0.72 |
| Pyramidal I <c+a> | $\xi_0$ (MPa) | 648 | 490 |
| | $\xi_\infty$ (MPa) | 1941 | 742 |
| | $h_0$ (MPa) | 5996 | 2689 |
| | a | 0.46 | 0.77 |

Since the Nelder-Mead method is a non-gradient-based local optimizer, the resulting optimal values are sensitive to the initial guess values. To investigate the influence of the initial guess values, a sensitivity analysis was conducted in which the initial guess value for each parameter was varied by +10 %, and the resulting change in optimal values of the model parameters was quantified and plotted in the form of a heatmap in Figure 13. Due to perturbations in the initial values, the percentage change in losses from the reference optimization run was less than 0.3% in all cases.

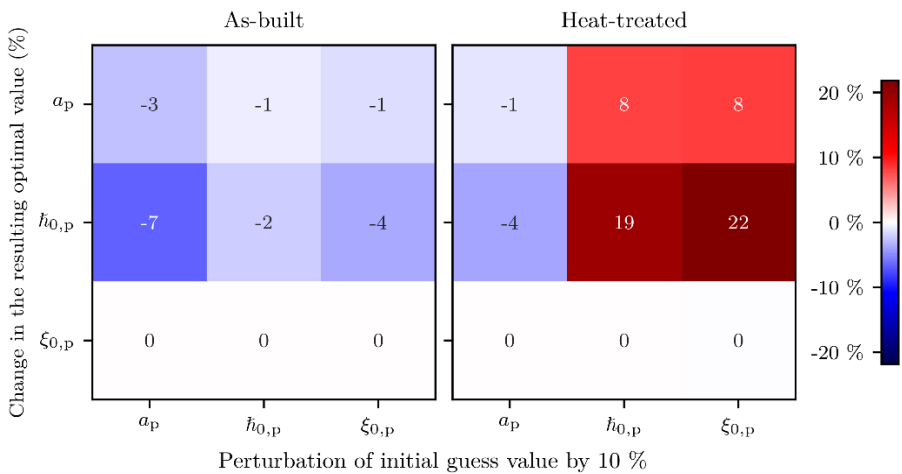

*Figure 13: A sensitivity study on the input initial guess value of the parameters of the "global" optimizer. The parameters on the x-axis are perturbed by 10 %, and the corresponding change (in %) of the optimal values from the reference is displayed on the y-axis. The relative change in losses from the reference was found to be less than 0.3 %.*

It is clear from both heatmaps that a 10 % perturbation in the initial guess for any of the three parameters has no influence on the optimal value of $\xi_{0,p}$. This indicates that the global response of the material is highly sensitive to $\xi_{0,p}$, and thus, the optimizer always arrives at a similar optimal value, irrespective of the initial guess values. Unlike $\xi_{0,p}$, the optimized values for $h_0$ and $a$, which jointly determine the shape of the hardening curve, exhibited dependence on the initial values. These deviations suggest that the system is relatively less sensitive to $a$ and $h_0$, with numerous parameter combinations producing a response similar to that of the optimized values. Sedighiani et al. investigated the sensitivities of the adjustable parameters in the phenomenological crystal plasticity model used in the present study and confirmed the current observations: the critical resistance to slip can be uniquely identified, whereas scatter exists in the optimal values of $a$ and $h_0$, with its magnitude increasing in that order [60].

The relative changes in the resultant optimized values are higher in the heat-treated specimen compared to the as-built specimen, attributable to the difference in slip system hardening. Here, the larger the magnitude of stress that a slip system is allowed to harden, the lower the percentage deviation from reference optimal values, as seen in Figure 13.

### 3.4.3. Deformation mechanisms

Figure 14 categorizes the grains based on the dominant slip family. Here, the dominant slip family is determined by identifying the slip family that accommodates the maximum plastic shear.

In both specimens, the deformation in the {10.0} grains is primarily accommodated by prismatic slip, although a few {10.0} grains in the heat-treated specimen deform via pyramidal I <c+a> slip. The absence of pyramidal I <c+a> slip in {10.0} grains of the as-built sample may be attributed to its higher CRSS. Next, the {10.2} grains primarily deform through the activation of the basal slip system. Whereas, the {10.1} grains deform via activation of both basal and prismatic slip systems. The deformation mechanisms of {00.2} grains differ significantly between the two samples. In the heat-treated sample, deformation in {00.2} grains is solely accommodated by pyramidal I <c+a> slip, as shown in Figure 14. Whereas, in the as-built specimen, numerous {00.2} grains also deform by the basal and prismatic slip in addition to the pyramidal I <c+a> slip system.

The lattice strain in {00.2} planes represents an average strain along the c-axis in the grains, having the c-axis parallel to the loading direction. As suggested by Figure 14, most of the diffracting {00.2} grains deform by pyramidal I <c+a> slip and rest either by basal or prismatic slip. One possible hypothesis to explain the {00.2} lattice strain evolution is as follows. Ideally, the lattice strain evolution of {00.2} grains deforming via pyramidal I <c+a> slip would plateau, similar to other grain families experiencing plastic deformation. However, {00.2} grains deforming via basal and prismatic slip would continue to deform elastically along the c-axis, and the strain evolution along the c-axis would never plateau, as <a> dislocations on the basal or prismatic plane would not cause plastic strain along the c-axis. As a result, when the contributions from both types of grains are combined, the overall lattice strain continues to increase. Moreover, the broadening of the {00.2} reflection can be attributed to the increased dislocation density in grains deforming via the pyramidal I <c+a> slip system.

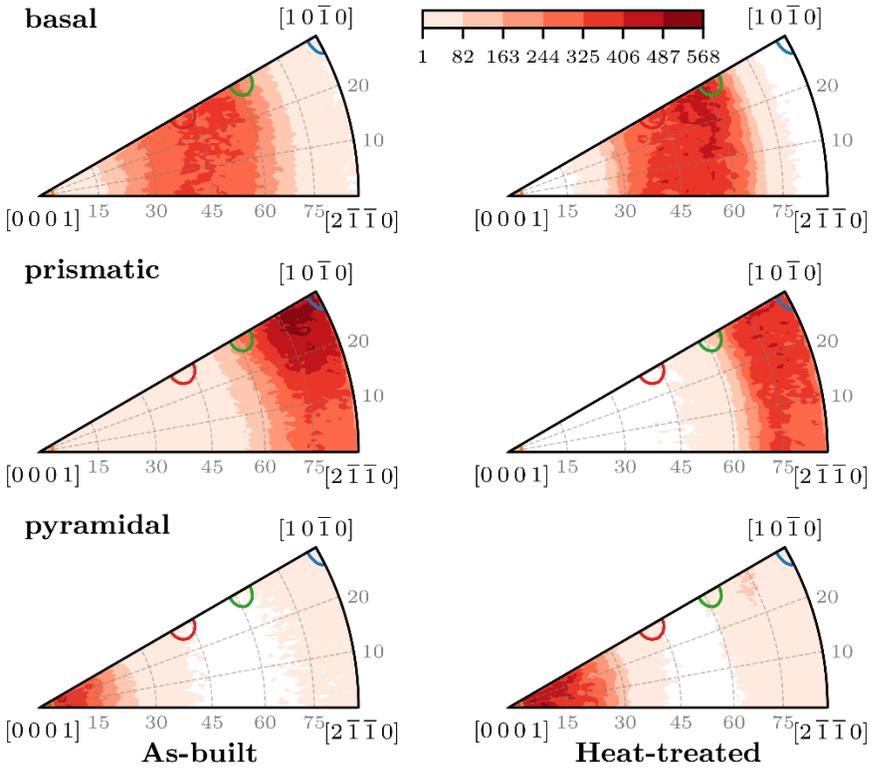

*Figure 14: Heatmaps showing the dominant deformation mode in different grains. The grains are categorized into three categories based on dominant deformation mechanisms – basal, prismatic, and pyramidal I <c+a>.*

### 3.4.4. Intensity evolution

Although diffraction peak intensities cannot be directly modeled with crystal plasticity simulations, the number of grains with specific orientations—proportional to the diffraction peak intensity—was tracked as a function of applied stress for both heat-treated and as-built conditions (see Figure 15). To this end, the number of grains belonging to the {10.0}, {00.2}, {10.1}, and {10.2} orientations in both the undeformed and deformed configurations was tracked as a function of applied stress for the heat-treated and as-built conditions. The model qualitatively captured the intensity evolution trend for {10.0}, {10.1}, and {10.2} reflections in both specimens. However, there was a significant discrepancy between the experimentally tracked intensity evolution and predicted intensity evolution trends for the {00.2} grains, particularly in the as-built state, and to a lesser extent, in the heat-treated condition. For the as-built state, the simulation predicted intensification of {00.2} reflection as opposed to the attenuation observed in the experiments. To further understand the intensity evolution trend, grain rotation tendencies were studied individually in each of the four grain families.

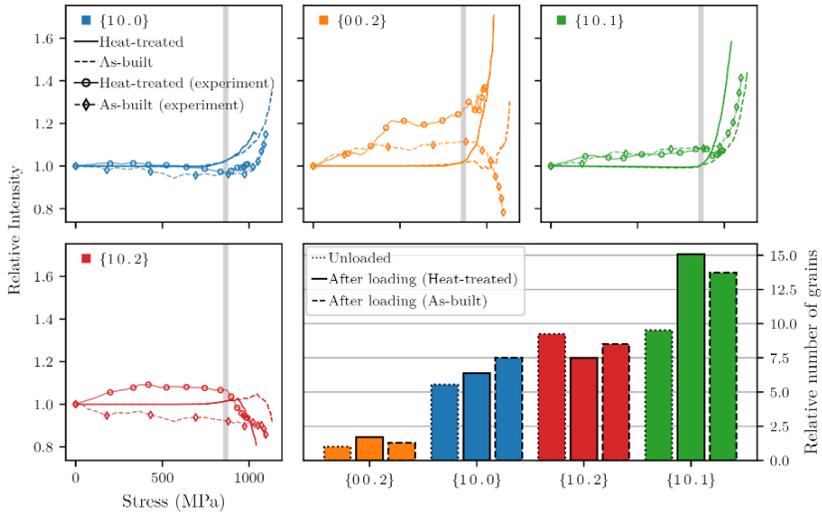

Figure 15: Comparison of the peak intensity evolution, determined from experiments, with the number of grains of corresponding orientations extracted from simulations. Both values have been normalized by the corresponding values at the undeformed configuration. The grey patches in the plots represent elastoplastic transition regions identified from global stress-strain data.

The grain rotation tendencies upon tensile deformation for the as-built and heat-treated specimens are shown in Figure 16. For each grain family, the center of mass is marked to track its rotation tendency. The poles of {71.0}, {21.0}, and {11.0} grains, upon deforming by prismatic slip, spread toward the [10-10] vertex of the IPF. This pole movement corresponds to a grain rotation along the c-axis, attempting to align the <10-10> direction with the loading axis. This corroborates the grain rotation trend previously associated with the prismatic slip, investigated by EBSD [55,61–63]. The spreading of the {11.0} poles due to prismatic slip activation explains the attenuation of the {11.0} reflection, while the rotation of nearby orientations, such as {71.0}, towards {10.0} can explain the intensification of the {10.0} reflection.

In both specimens, the {10.1} and {10.2} grains were observed to move away from the [0001] vertex in the IPF, indicating an increased angle of inclination between the loading direction and the c-axis. This observation is consistent with the findings of Lin et al., who, through EBSD analysis, demonstrated that basal slip causes grain rotation in a way that tends to align the basal plane with the loading direction [63]. Ideally, this rotation would have resulted in reduced intensity for both {10.1} and {10.2} reflections as these grains move out of the diffraction conditions. However, as shown in Figures 8(D) and 11(D), {10.2} reflection attenuates after deformation, while {10.1} reflection intensifies. To better understand this trend, the evolution of neighboring grain orientations, such as {10.3}, {10.4}, and {71.10}, was also tracked. It was observed that basal slip activation in the {71.10} and {10.3} grains rotated them into an orientation that enhances the intensity of the {10.2} reflection. Similarly, basal slip activation in {71.10} grains rotated them into an orientation that contributes to the {10.1} diffraction peak. Therefore, the post-deformation intensities of {10.1} and {10.2} depend on the relative balance of grains entering and leaving the respective diffraction condition. As a result, the intensification of the {10.1} reflection and the attenuation of the {10.2} reflection is not typical of basal slip, and intensity variation for {10.1} and {10.2} depends upon the initial texture.

Finally, {00.2} grains tend to deform via pyramidal I <c+a> slip system in both samples, as seen from Figure 13. However, pyramidal I <c+a> slip activation did not cause significant grain rotation in the {00.2} grains. Hence, the rotation of {10.10} grains was studied to understand the rotation induced by pyramidal I <c+a> slip. In the heat-treated specimen, {10.10} grains, upon activation of pyramidal I <c+a> slip, consistently moved towards the [0001] vertex, meaning that pyramidal I <c+a> slip rotates a grain in a way that aligns the c-axis with the loading axis. This brings orientation such as {10.10} into the {00.2} diffraction condition, resulting in increased intensity of {00.2} reflection in the heat-treated state (Figure 11 (D)). However, such coordinated movement of {10.10} grains towards the [0001] vertex was not present for the as-built sample. This could be because of {10.10} grains that deform by basal or prismatic slip, instead of pyramidal I <c+a> slip, would not rotate toward the [0001] corner. Moreover, some of the {00.2} grains that deform by basal or prismatic slip would also move out of the {00.2} diffraction condition. In this way, the drop in {00.2} peak intensity after yielding in the as-built specimen can be explained.

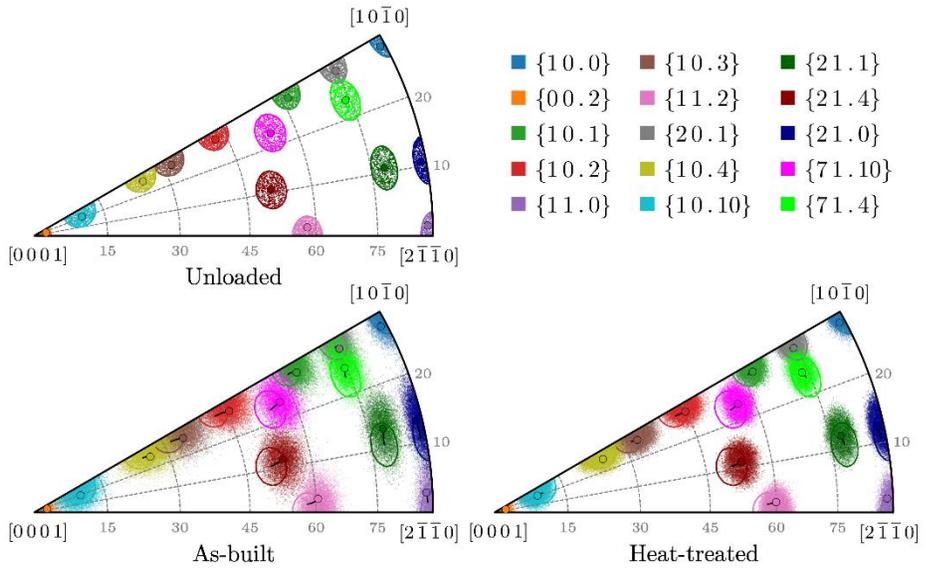

*Figure 16: (A) The IPF triangle indicates the initial location for 15 different grain families that were monitored during the simulations in an undeformed state. The grain rotation tendencies for 15 different grain families during the tensile deformation in (B) heat-treated and (C) as-built conditions are shown. Bold lines demarcate a threshold angle of 3.5 degrees, which was chosen as the tolerance to determine grains that satisfy the diffraction condition. The markers with black colored edges represent the mean of the coordinates in the IPF triangle, and the black line represents the path traversed during loading.*

## 3.5. Possible factors affecting the CRSS of the two hexagonal phases (α′ and α)

This section primarily discusses the potential origin for the observed differences in CRSS values of the different slip systems in the two hexagonal phases (α and α′): the c/a ratio of the hexagonal unit cell, the chemical composition of the hexagonal unit cell, the effective grain sizes, and the heterogeneous distribution of intergranular stresses in the α′ phase.

*(I) c/a ratio*

The c/a ratio of a hexagonal unit cell is a crucial geometric factor influencing the activity of slip systems. A higher c/a ratio favors basal slip activation, while a lower c/a ratio predisposes the hexagonal unit cell to deformation via non-basal slip systems, viz., prismatic <a>, pyramidal I <c+a>, or pyramidal II <c+a> [51,64]. However, both phases have a similar c/a ratio (see Table 4), meaning the c/a ratio cannot account for the differences in CRSS between the two phases.

*Table 4: Unit cell parameters for α and α' phases*

| Phase | a (nm) | c (nm) | c/a |
|-------|--------|--------|--------|
| α     | 0.292  | 0.467  | 1.5969 |
| α'    | 0.294  | 0.468  | 1.5952 |

*(II) Chemical composition*

EDS analysis revealed a higher Al content in the α phase (≈ 7 wt.%) compared to the α' phase (6 wt.%), whereas the V content showed the opposite trend. Generally, an increased Al content strengthens the α phase; however, Garg et al., based on ab-initio simulations, demonstrated that this effect is highly anisotropic [65]. As such, the α phase would be expected to exhibit higher CRSS values across all slip systems than the α' phase, though the magnitude of the increment would differ for each slip system. However, contrary to this expectation, only the prismatic slip system in the α phase exhibits a higher CRSS, while the basal and pyramidal slip systems have lower CRSS values.

*(III) Effective grain size*

The α and α' phases differ significantly in grain size. An inverse relationship between CRSS values and grain size has been previously observed in both pure titanium and Ti-6Al-4V alloys [29,54,66]. The much thinner α' laths, compared to the α grains, would have resulted in higher CRSS values for the α' phase. Surprisingly, this expected trend was not reflected in the CRSS values obtained from the optimization procedure.

*(IV) Heterogeneous distribution of intergranular stresses in α' phase*

Macroscopic residual stresses (Type I residual stresses), induced by L-PBF processing, in the α' phase have been extensively reported in the literature [67–69]. However, the distribution of Type II intergranular stresses across different orientations has never been explored in the as-built Ti-6Al-4V. Such stresses could result from thermal and mechanical anisotropy or from α' transformation or from a combination of both. Although, not in Ti-6Al-4V, the thermally induced residual stresses have been observed in hexagonal unit cells of beryllium [70,71]. Similarly, the presence of α' transformation-induced stresses during rapid cooling has been confirmed in steels but not in titanium-based alloys [72,73].

If present, α′ transformation-induced and thermally induced stresses in the α′ phase would influence the onset of plasticity in different grain families. Specifically, tensile residual stresses might hasten the initiation of plastic deformation, whereas compressive residual stresses could delay it. Consequently, neglecting these residual stresses during crystal plasticity simulations could lead to an underestimation or overestimation of CRSS values, depending on their tensile or compressive nature, respectively. This could explain the wide disparity in CRSS values reported for the α phase in Ti-6Al-4V. For instance, the ratio of $\xi_b/\xi_p$ has been reported to range from 0.9 to 1.45, and an even greater variance has been observed for the ratio of $\xi_b/\xi_\pi$, which ranges from 1.5 to 5 [74–76].

## 4. Summary

This study investigated and compared the micromechanical deformation behavior of the α′ and α phases in laser powder bed fusion (L-PBF) processed Ti-6Al-4V, in the as-built and heat-treated conditions, respectively, using an integrated experimental and simulation approach.

The distinct lattice strain evolutions observed in various diffracting planes for the α′ and α phases highlighted their inherent elastic and plastic anisotropy. Both phases exhibited qualitatively similar anisotropic elastic behavior, with the highest elastic modulus along the {00.2} planes and the lowest along the {10.0} planes—confirming the previously reported inverse correlation between apparent Young's modulus and the angle between the c-axis and the loading direction. Quantitatively, however, the α phase consistently exhibited a higher elastic modulus in all directions compared to the α′ phase. However, the two phases (α′ and α) displayed markedly different plastic deformation responses, reflected in the differing slip activation sequences. In the α′ phase, the sequence was prismatic → basal → pyramidal I <c+a>, whereas in the α phase, it followed a basal → prismatic → pyramidal I <c+a> order. Additionally, changes in full width at half maximum (FWHM) and diffraction peak intensity for various reflections were analyzed to further elucidate the plastic deformation behavior.

Next, an inverse modeling approach based on the Nelder-Mead optimization algorithm was implemented within a crystal plasticity framework. This approach utilized both the global tensile response and local lattice strain–stress data for four reflections—{10.0}, {00.2}, {10.1}, and {10.2}—to determine the elastic constants and slip system-specific critical resolved shear stresses (CRSS) for the α′ and α phases. The results revealed that the α′ phase exhibited higher CRSS values for all slip families except the prismatic slip family. Potential explanations for these differences were discussed, with intergranular stresses present in the undeformed state suggested as a contributing factor. Furthermore, crystal plasticity simulations using the optimized parameter set provided insight into the deformation mechanisms across different grain orientations, collectively explaining the evolution of both lattice strains and diffraction peak intensities.

## Acknowledgement

The work leading to this publication has been funded by the ICON project "PROCSIMA", which fits in the MacroModelMat (M3) research program, coordinated by Siemens (Siemens Digital Industries Software, Belgium) and funded by SIM (Strategic Initiative Materials in Flanders) and VLAIO (Flanders Innovation and Entrepreneurship).The computational resources and services used in this work were provided by the VSC (Flemish Supercomputer Center), funded by the Research Foundation—Flanders (FWO) and the Flemish Government.

We acknowledge DESY (Hamburg, Germany), a member of the Helmholtz Association HGF, for the provision of experimental facilities. Parts of this research were carried out at PETRA III and we would like to thank Dr. Andreas Stark, Dr. Emad Maawad and Dr. Norbert Schell for assistance in using beamline P07 and the in-situ dilatometer. Beamtime was allocated for proposal I-20190769 EC. This work has further benefitted from support by the project CALIPSOplus [Grant Agreement 730872] from the EU Framework Programme for Research and Innovation HORIZON 2020.

## Data Availability

All the codes and data used during this study are available at https://doi.org/10.5281/zenodo.15357915

# 5. References

[1] 3DSystems. https://www.3dsystems.com/materials (accessed October 26, 2024).

[2] E. Marin, A. Lanzutti, Biomedical Applications of Titanium Alloys: A Comprehensive Review, Materials 17 (2024). https://doi.org/10.3390/ma17010114

[3] S. Cao, Y. Zou, C.V.S. Lim, X. Wu, Review of laser powder bed fusion (LPBF) fabricated Ti-6Al-4V: process, post-process treatment, microstructure, and property, Light: Advanced Manufacturing 2 (2021). https://doi.org/10.37188/lam.2021.020

[4] S.D. Jadhav, Laser-Based Powder Bed Fusion Additive Manufacturing of Highly Conductive Copper and Copper Alloys, PhD Thesis, KU Leuven, 2021. https://lirias.kuleuven.be/3613387 (accessed February 4, 2025).

[5] L. Thijs, Microstructure and texture of metal parts produced by Selective Laser Melting, PhD Thesis, KU Leuven, 2014. https://lirias.kuleuven.be/1748330 (accessed February 4, 2025).

[6] P. Mercelis, Control of selective laser sintering and selective laser melting processes, PhD Thesis, KU Leuven, 2007. https://lirias.kuleuven.be/1747308 (accessed February 4, 2025).

[7] P. Barriobero-Vila, K. Artzt, A. Stark, N. Schell, M. Siggel, J. Gussone, J. Kleinert, W. Kitsche, G. Requena, J. Haubrich, Mapping the geometry of Ti-6Al-4V: From martensite decomposition to localized spheroidization during selective laser melting, Scr Mater 182 (2020) 48–52. https://doi.org/10.1016/j.scriptamat.2020.02.043

[8] B. Vrancken, Study of Residual Stresses in Selective Laser Melting, PhD Thesis, KU Leuven, 2016. https://lirias.kuleuven.be/1942277 (accessed February 4, 2025).

[9] B. Vrancken, L. Thijs, J.P. Kruth, J. Van Humbeeck, Heat treatment of Ti6Al4V produced by Selective Laser Melting: Microstructure and

mechanical properties, J Alloys Compd 541 (2012) 177–185. https://doi.org/10.1016/j.jallcom.2012.07.022

[10] C. V. Funch, A. Palmas, K. Somlo, E.H. Valente, X. Cheng, K. Poulios, M. Villa, M.A.J. Somers, T.L. Christiansen, Targeted heat treatment of additively manufactured Ti-6Al-4V for controlled formation of Bi-lamellar microstructures, J Mater Sci Technol 81 (2021) 67–76. https://doi.org/10.1016/j.jmst.2021.01.004

[11] M.T. Tsai, Y.W. Chen, C.Y. Chao, J.S.C. Jang, C.C. Tsai, Y.L. Su, C.N. Kuo, Heat-treatment effects on mechanical properties and microstructure evolution of Ti-6Al-4V alloy fabricated by laser powder bed fusion, J Alloys Compd 816 (2020) 152615. https://doi.org/10.1016/j.jallcom.2019.152615

[12] P. Van Cauwenbergh, Tailoring heat treatments for metals processed by Laser Powder Bed Fusion, PhD Thesis, KU Leuven, 2022. https://lirias.kuleuven.be/3675190 (accessed February 4, 2025).

[13] R. Sabban, S. Bahl, K. Chatterjee, S. Suwas, Globularization using heat treatment in additively manufactured Ti-6Al-4V for high strength and toughness, Acta Mater 162 (2019) 239–254. https://doi.org/10.1016/j.actamat.2018.09.064

[14] J. Su, F. Jiang, J. Li, C. Tan, Z. Xu, H. Xie, J. Liu, J. Tang, D. Fu, H. Zhang, J. Teng, Phase transformation mechanisms, microstructural characteristics and mechanical performances of an additively manufactured Ti-6Al-4V alloy under dual-stage heat treatment, Mater Des 223 (2022) 111240. https://doi.org/10.1016/j.matdes.2022.111240

[15] G.M. Ter Haar, T.H. Becker, Selective laser melting produced Ti-6Al-4V: Post-process heat treatments to achieve superior tensile properties, Materials 11 (2018). https://doi.org/10.3390/ma11010146

[16] S. Hémery, P. Villechaise, On the influence of ageing on the onset of plastic slip in Ti-6Al-4V at room temperature: Insight on dwell fatigue behavior, Scr Mater 130 (2017) 157–160. https://doi.org/10.1016/j.scriptamat.2016.11.042

[17]  J.A. Medina Perillaa, J. Gil Sevillano, Two-dimensional sections of the yield locus of a Ti-6%Al-4%V alloy with a strong transverse-type crystallographic α-texture, 1995. https://doi.org/10.1016/0921-5093(95)09780-5

[18]  F. Bridier, D.L. McDowell, P. Villechaise, J. Mendez, Crystal plasticity modeling of slip activity in Ti-6Al-4V under high cycle fatigue loading, Int J Plast 25 (2009) 1066–1082. https://doi.org/10.1016/j.ijplas.2008.08.004

[19]  I.P. Jones, W.B. Hutchinson, Stress-state dependence of slip in Ti-6Al-4V and other H.C.P. metals, Acta Metallurgica 29 (1981) 951–968. https://doi.org/10.1016/0001-6160(81)90049-3

[20]  X. Song, S.Y. Zhang, D. Dini, A.M. Korsunsky, Finite element modelling and diffraction measurement of elastic strains during tensile deformation of HCP polycrystals, Comput Mater Sci 44 (2008) 131–137. https://doi.org/10.1016/j.commatsci.2008.01.043

[21]  A. Kurdi, A.K. Basak, Micro-mechanical behaviour of selective laser melted Ti6Al4V under compression, Materials Science and Engineering: A 826 (2021). https://doi.org/10.1016/j.msea.2021.141975

[22]  A. Orozco-Caballero, D. Lunt, J.D. Robson, J. Quinta da Fonseca, How magnesium accommodates local deformation incompatibility: A high-resolution digital image correlation study, Acta Mater 133 (2017) 367–379. https://doi.org/10.1016/j.actamat.2017.05.040

[23]  A. Zeghadi, S. Forest, A.F. Gourgues, O. Bouaziz, Ensemble averaging stress-strain fields in polycrystalline aggregates with a constrained surface microstructure-part 2: Crystal plasticity, in: Philosophical Magazine, Taylor and Francis Ltd., 2007: pp. 1425–1446. https://doi.org/10.1080/14786430601009517

[24]  A. Zeghadi, F. N'Guyen, S. Forest, A.F. Gourgues, O. Bouaziz, Ensemble averaging stress-strain fields in polycrystalline aggregates with a constrained surface microstructure-part 1: Anisotropic elastic behaviour, in: Philosophical Magazine, Taylor and Francis Ltd., 2007: pp. 1401–1424. https://doi.org/10.1080/14786430601009509

[25] M. Diehl, P. Shanthraj, P. Eisenlohr, F. Roters, Neighborhood influences on stress and strain partitioning in dual-phase microstructures: An investigation on synthetic polycrystals with a robust spectral-based numerical method, Meccanica 51 (2016) 429–441. https://doi.org/10.1007/s11012-015-0281-2

[26] Y. Peng, K. Miao, W. Sun, C. Liu, H. Wu, L. Geng, G. Fan, Recent Progress of Synchrotron X-Ray Imaging and Diffraction on the Solidification and Deformation Behavior of Metallic Materials, Acta Metallurgica Sinica (English Letters) 35 (2022) 3–24. https://doi.org/10.1007/s40195-021-01311-4

[27] P.P. Dhekne, M. Bonisch, M. Seefeldt, K. Vanmeensel, In-situ synchrotron X-ray diffraction investigation of martensite decomposition in Laser Powder Bed Fusion ( L-PBF ) processed Ti – 6Al – 4V, Material Science and Engineering A 899 (2024). https://doi.org/10.1016/j.msea.2024.146421

[28] P. Barnes, The use of synchrotron diffraction for studying energy-dispersive chemical reactions, Journal of Physics and Chemistry of Solids 52 (1991) 1299–1306. https://doi.org/10.1016/0022-3697(91)90207-G

[29] L. Wang, Z. Zheng, H. Phukan, P. Kenesei, J.S. Park, J. Lind, R.M. Suter, T.R. Bieler, Direct measurement of critical resolved shear stress of prismatic and basal slip in polycrystalline Ti using high energy X-ray diffraction microscopy, Acta Mater 132 (2017) 598–610. https://doi.org/10.1016/j.actamat.2017.05.015

[30] D. Zhang, L. Wang, H. Zhang, A. Maldar, G. Zhu, W. Chen, J.S. Park, J. Wang, X. Zeng, Effect of heat treatment on the tensile behavior of selective laser melted Ti-6Al-4V by in situ X-ray characterization, Acta Mater 189 (2020) 93–104. https://doi.org/10.1016/j.actamat.2020.03.003

[31] J.M. Vallejos, P. Barriobero-Vila, J. Gussone, J. Haubrich, K. Kelm, A. Stark, N. Schell, G. Requena, In Situ High-Energy Synchrotron X-Ray Diffraction Reveals the Role of Texture on the Activation of Slip and

Twinning during Deformation of Laser Powder Bed Fusion Ti–6Al–4V, Adv Eng Mater 23 (2021). https://doi.org/10.1002/adem.202001556

[32] T. Ungár, Dislocation densities, arrangements and character from X-ray diffraction experiments, 2001. https://doi.org/10.1016/S0921-5093(00)01685-3

[33] Z.Y. Zhong, H.G. Brokmeier, W.M. Gan, E. Maawad, B. Schwebke, N. Schell, Dislocation density evolution of AA 7020-T6 investigated by in-situ synchrotron diffraction under tensile load, Mater Charact 108 (2015) 124–131. https://doi.org/10.1016/j.matchar.2015.09.004

[34] Z.H. Cong, N. Jia, X. Sun, Y. Ren, J. Almer, Y.D. Wang, Stress and strain partitioning of ferrite and martensite during deformation, Metall Mater Trans A Phys Metall Mater Sci 40 (2009) 1383–1387. https://doi.org/10.1007/s11661-009-9824-2

[35] H. Zhang, A. Jérusalem, E. Salvati, C. Papadaki, K.S. Fong, X. Song, A.M. Korsunsky, Multi-scale mechanisms of twinning-detwinning in magnesium alloy AZ31B simulated by crystal plasticity modeling and validated via in situ synchrotron XRD and in situ SEM-EBSD, Int J Plast 119 (2019) 43–56. https://doi.org/10.1016/j.ijplas.2019.02.018

[36] MAUD, (2011). http://maud.radiographema.com/ (accessed November 14, 2024).

[37] S. Matthies, H.R. Wenk, G.W. Vinel, Some basic concepts of texture analysis and comparison of three methods to calculate orientation distributions from pole figures, J Appl Crystallogr 21 (1988) 285–304. https://doi.org/10.1107/S0021889888000275

[38] A.K. Singh, H.K. Mao, J. Shu, R.J. Hemley, Estimation of single-crystal elastic moduli from polycrystalline x-ray diffraction at high pressure: Application to feo and iron, Phys Rev Lett 80 (1998) 2157–2160. https://doi.org/10.1103/PhysRevLett.80.2157

[39] A.K. Singh, C. Balasingh, The lattice strains in a specimen (cubic system) compressed nonhydrostatically in an opposed anvil high pressure setup, J Appl Phys 75 (1994) 4956–4962. https://doi.org/10.1063/1.355786

[40] F. Roters, M. Diehl, P. Shanthraj, P. Eisenlohr, C. Reuber, S.L. Wong, T. Maiti, A. Ebrahimi, T. Hochrainer, H.O. Fabritius, S. Nikolov, M. Friák, N. Fujita, N. Grilli, K.G.F. Janssens, N. Jia, P.J.J. Kok, D. Ma, F. Meier, E. Werner, M. Stricker, D. Weygand, D. Raabe, DAMASK – The Düsseldorf Advanced Material Simulation Kit for modeling multi-physics crystal plasticity, thermal, and damage phenomena from the single crystal up to the component scale, Comput Mater Sci 158 (2019) 420–478. https://doi.org/10.1016/j.commatsci.2018.04.030

[41] J.W. Hutchinson, Elastic-plastic behavior of polycrystalline metals and composites, Proceedings of the Royal Society of London A 319 (1970) 247–272. https://doi.org/10.1098/rspa.1970.0177

[42] J.A. Nelder, R. Mead, A Simplex Method for function minimization, Comput J 8 (1965). https://doi.org/https://doi.org/10.1093/comjnl/8.1.27

[43] P. Virtanen, R. Gommers, T.E. Oliphant, M. Haberland, T. Reddy, D. Cournapeau, E. Burovski, P. Peterson, W. Weckesser, J. Bright, S.J. van der Walt, M. Brett, J. Wilson, K.J. Millman, N. Mayorov, A.R.J. Nelson, E. Jones, R. Kern, E. Larson, C.J. Carey, İ. Polat, Y. Feng, E.W. Moore, J. VanderPlas, D. Laxalde, J. Perktold, R. Cimrman, I. Henriksen, E.A. Quintero, C.R. Harris, A.M. Archibald, A.H. Ribeiro, F. Pedregosa, P. van Mulbregt, A. Vijaykumar, A. Pietro Bardelli, A. Rothberg, A. Hilboll, A. Kloeckner, A. Scopatz, A. Lee, A. Rokem, C.N. Woods, C. Fulton, C. Masson, C. Häggström, C. Fitzgerald, D.A. Nicholson, D.R. Hagen, D. V. Pasechnik, E. Olivetti, E. Martin, E. Wieser, F. Silva, F. Lenders, F. Wilhelm, G. Young, G.A. Price, G.L. Ingold, G.E. Allen, G.R. Lee, H. Audren, I. Probst, J.P. Dietrich, J. Silterra, J.T. Webber, J. Slavič, J. Nothman, J. Buchner, J. Kulick, J.L. Schönberger, J.V. de Miranda Cardoso, J. Reimer, J. Harrington, J.L.C. Rodríguez, J. Nunez-Iglesias, J. Kuczynski, K. Tritz, M. Thoma, M. Newville, M. Kümmerer, M. Bolingbroke, M. Tartre, M. Pak, N.J. Smith, N. Nowaczyk, N. Shebanov, O. Pavlyk, P.A. Brodtkorb, P. Lee, R.T. McGibbon, R. Feldbauer, S. Lewis, S. Tygier, S. Sievert, S. Vigna, S. Peterson, S. More, T. Pudlik, T. Oshima, T.J. Pingel, T.P. Robitaille, T. Spura, T.R. Jones, T. Cera, T. Leslie, T. Zito, T. Krauss, U. Upadhyay, Y.O. Halchenko, Y. Vázquez-Baeza, SciPy 1.0: fundamental algorithms for scientific computing in

Python, Nat Methods 17 (2020) 261–272. https://doi.org/10.1038/s41592-019-0686-2

[44] E. Wielewski, D.E. Boyce, J.S. Park, M.P. Miller, P.R. Dawson, A methodology to determine the elastic moduli of crystals by matching experimental and simulated lattice strain pole figures using discrete harmonics, Acta Mater 126 (2017) 469–480. https://doi.org/10.1016/j.actamat.2016.12.026

[45] T. Dick, G. Cailletaud, Fretting modelling with a crystal plasticity model of Ti6Al4V, Comput Mater Sci 38 (2006) 113–125. https://doi.org/10.1016/j.commatsci.2006.01.015

[46] D. Dunst, H. Mecking, Analysis of experimental and theoretical rolling textures of two-phase titanium alloys, International Journal of Materials Research 87 (1996). https://doi.org/10.1515/ijmr-1996-870613

[47] G. Lutjering, J.C. Williams, Titanium, 2nd ed., Springer Berlin, Heidelberg, 2007. https://doi.org/10.1007/978-3-540-73036-1

[48] P. Erdely, P. Staron, E. Maawad, N. Schell, H. Clemens, S. Mayer, Lattice and phase strain evolution during tensile loading of an intermetallic, multi-phase γ-TiAl based alloy, Acta Mater 158 (2018) 193–205. https://doi.org/10.1016/j.actamat.2018.07.062

[49] D. Gloaguen, B. Girault, J. Fajoui, V. Klosek, M.J. Moya, In situ lattice strains analysis in titanium during a uniaxial tensile test, Materials Science and Engineering: A 662 (2016) 395–403. https://doi.org/10.1016/j.msea.2016.03.089

[50] D. Gloaguen, G. Oum, V. Legrand, J. Fajoui, M.J. Moya, T. Pirling, W. Kockelmann, Intergranular Strain Evolution in Titanium During Tensile Loading: Neutron Diffraction and Polycrystalline Model, Metall Mater Trans A Phys Metall Mater Sci 46 (2015) 5038–5046. https://doi.org/10.1007/s11661-015-3073-3

[51] K. Sofinowski, M. Šmíd, S. van Petegem, S. Rahimi, T. Connolley, H. van Swygenhoven, In situ characterization of work hardening and

springback in grade 2 α-titanium under tensile load, Acta Mater 181 (2019) 87–98. https://doi.org/10.1016/j.actamat.2019.09.039

[52] K.E. Agbovi, B. Girault, J. Fajoui, S. Kabra, W. Kockelmann, T. Buslaps, A. Poulain, D. Gloaguen, Lattice strain development in an alpha titanium alloy studied using synchrotron and neutron diffraction, Materials Science and Engineering: A 819 (2021). https://doi.org/10.1016/j.msea.2021.141489

[53] I.C. Dragomir, T. Ungár, Contrast factors of dislocations in the hexagonal crystal system, J Appl Crystallogr 35 (2002) 556–564. https://doi.org/10.1107/S0021889802009536

[54] E. Sjögren-Levin, W. Pantleon, A. Ahadi, Z. Hegedüs, U. Lienert, N. Tsuji, K. Ameyama, D. Orlov, Grain-level mechanism of plastic deformation in harmonic structure materials revealed by high resolution X-ray diffraction, Acta Mater 265 (2024). https://doi.org/10.1016/j.actamat.2023.119623

[55] P.C. Zhao, B. Chen, Z.G. Zheng, B. Guan, X.C. Zhang, S.T. Tu, Microstructure and Texture Evolution in a Post-dynamic Recrystallized Titanium During Annealing, Monotonic and Cyclic Loading, Metall Mater Trans A Phys Metall Mater Sci 52 (2021) 394–412. https://doi.org/10.1007/s11661-020-06071-x

[56] A. Huet, A. Naït-Ali, T. Giroud, P. Villechaise, S. Hémery, Onset of plastic deformation and strain localization in relation to β phase in metastable β and dual phase Ti alloys, Acta Mater 240 (2022) 118348. https://doi.org/10.1016/j.actamat.2022.118348

[57] D. Gloaguen, G. Oum, V. Legrand, J. Fajoui, S. Branchu, Experimental and theoretical studies of intergranular strain in an alpha titanium alloy during plastic deformation, Acta Mater 61 (2013) 5779–5790. https://doi.org/10.1016/j.actamat.2013.06.022

[58] J. Čapek, E. Polatidis, M. Knapek, C. Lyphout, N. Casati, R. Pederson, M. Strobl, The Effect of γ″ and δ Phase Precipitation on the Mechanical Properties of Inconel 718 Manufactured by Selective Laser Melting: An In Situ Neutron Diffraction and Acoustic Emission

Study, Jom 73 (2021) 223–232. https://doi.org/10.1007/s11837-020-04463-3

[59] A. Borbély, The modified Williamson-Hall plot and dislocation density evaluation from diffraction peaks, Scr Mater 217 (2022). https://doi.org/10.1016/j.scriptamat.2022.114768

[60] K. Sedighiani, M. Diehl, K. Traka, F. Roters, J. Sietsma, D. Raabe, An efficient and robust approach to determine material parameters of crystal plasticity constitutive laws from macro-scale stress-strain curves, Int J Plast 134 (2020). https://doi.org/10.1016/j.ijplas.2020.102779

[61] A. Chapuis, Q. Liu, Simulations of texture evolution for HCP metals: Influence of the main slip systems, Comput Mater Sci 97 (2015) 121–126. https://doi.org/10.1016/j.commatsci.2014.10.017

[62] Z. Huang, L. Wang, B. Zhou, T. Fischer, S. Yi, X. Zeng, Observation of non-basal slip in Mg-Y by in situ three-dimensional X-ray diffraction, Scr Mater 143 (2018) 44–48. https://doi.org/10.1016/j.scriptamat.2017.09.011

[63] X. Lin, Z. Chen, J. Shao, J. Xiong, Z. Hu, C. Liu, Deformation mechanism, orientation evolution and mechanical properties of annealed cross-rolled Mg-Zn-Zr-Y-Gd sheet during tension, Journal of Magnesium and Alloys 11 (2023) 2340–2350. https://doi.org/10.1016/j.jma.2021.08.006

[64] H. Li, D.E. Mason, T.R. Bieler, C.J. Boehlert, M.A. Crimp, Methodology for estimating the critical resolved shear stress ratios of α-phase Ti using EBSD-based trace analysis, Acta Mater 61 (2013) 7555–7567. https://doi.org/10.1016/j.actamat.2013.08.042

[65] P. Garg, M.A. Bhatia, K.N. Solanki, Uncovering the influence of metallic and non-metallic impurities on the ideal shear strength and ductility of Ti: An ab-initio study, J Alloys Compd 788 (2019) 413–421. https://doi.org/10.1016/j.jallcom.2019.02.231

[66] F. Benmessaoud, M. Cheikh, V. Velay, V. Vidal, H. Matsumoto, Role of grain size and crystallographic texture on tensile behavior induced by

sliding mechanism in Ti-6Al-4V alloy, Materials Science and Engineering: A 774 (2020) 138835. https://doi.org/10.1016/j.msea.2019.138835

[67] D. Gloaguen, B. Girault, B. Courant, P.A. Dubos, M.J. Moya, F. Edy, J. Rebelo Kornmeier, Study of Residual Stresses in Additively Manufactured Ti-6Al-4V by Neutron Diffraction Measurements, Metall Mater Trans A Phys Metall Mater Sci 51 (2020) 951–961. https://doi.org/10.1007/s11661-019-05538-w

[68] N.C. Levkulich, S.L. Semiatin, J.E. Gockel, J.R. Middendorf, A.T. DeWald, N.W. Klingbeil, The effect of process parameters on residual stress evolution and distortion in the laser powder bed fusion of Ti-6Al-4V, Addit Manuf 28 (2019) 475–484. https://doi.org/10.1016/j.addma.2019.05.015

[69] K. Artzt, T. Mishurova, P.P. Bauer, J. Gussone, P. Barriobero-Vila, S. Evsevleev, G. Bruno, G. Requena, J. Haubrich, Pandora's box-influence of contour parameters on roughness and subsurface residual stresses in laser powder bed fusion of Ti-6Al-4V, Materials 13 (2020) 1–24. https://doi.org/10.3390/ma13153348

[70] D.W. Brown, T.A. Sisneros, B. Clausen, S. Abeln, M.A.M. Bourke, B.G. Smith, M.L. Steinzig, C.N. Tomé, S.C. Vogel, Development of intergranular thermal residual stresses in beryllium during cooling from processing temperatures, Acta Mater 57 (2009) 972–979. https://doi.org/10.1016/j.actamat.2008.09.044

[71] C. Zhang, H. Li, B. Pang, H. Wang, J. Li, Z. Yang, G. Sun, Quantifying the inter- and intra-granular stresses and their effects on deformation behaviors for the polycrystalline hexagonal beryllium by neutron diffraction, Mater Charact 145 (2018) 534–544. https://doi.org/10.1016/j.matchar.2018.09.014

[72] D. Fukui, N. Nakada, S. Onaka, Internal residual stress originated from Bain strain and its effect on hardness in Fe–Ni martensite, Acta Mater 196 (2020) 660–668. https://doi.org/10.1016/j.actamat.2020.07.013

[73] D. Fukui, Y. Kawahito, N. Miyazawa, N. Nakada, Anisotropic cleavage fracture caused by transformation-induced internal stress in an as-

quenched martensite, Mater Charact 191 (2022) 112157. https://doi.org/10.1016/j.matchar.2022.112157

[74]  J.A. Medina Perilla, J. Gil Sevillano, Two-dimensional sections of the yield locus of a Ti-6Al-4V alloy with a strong transverse-type crystallographic α-texture, Materials Science and Engineering A 201 (1995) 103–110. https://doi.org/10.1016/0921-5093(95)09780-5

[75]  J.J. Fundenberger, M.J. Philippe, F. Wagner, C. Esling, Modelling and prediction of mechanical properties for materials with hexagonal symmetry (zinc, titanium and zirconium alloys), Acta Mater 45 (1997) 4041–4055. https://doi.org/10.1016/S1359-6454(97)00099-2

[76]  S.L. Semiatin, T.R. Bieler, Effect of texture and slip mode on the anisotropy of plastic flow and flow softening during hot working of Ti-6Al-4V, Metallurgical and Materials Transactions A 32 (2001). https://doi.org/10.1007/s11661-001-0155-1

Appendix A

*Table A1: Schmid factors for the basal <a> , prismatic <a>, and pyramidal I <c+a> slip system for nine α grain families*

| Slip System | {10.0} | {00.2} | {10.1} | {10.2} | {11.0} | {10.3} | {11.2} | {20.1} | {10.4} |
|---|---|---|---|---|---|---|---|---|---|
| **Basal <a> Slip System** | | | | | | | | | |
| (0002) [-12-10] | 0.000 | 0.000 | 0.000 | 0.000 | 0.000 | 0.000 | 0.225 | 0.000 | 0.000 |
| (0002) [-1-120] | 0.000 | 0.000 | 0.363 | 0.432 | 0.000 | 0.387 | 0.450 | 0.218 | 0.330 |
| (0002) [2-1-10] | 0.000 | 0.000 | 0.363 | 0.432 | 0.000 | 0.387 | 0.225 | 0.218 | 0.330 |
| **Prismatic <a> Slip System** | | | | | | | | | |
| (01-10) [2-110] | 0.433 | 0.000 | 0.335 | 0.199 | 0.433 | 0.119 | 0.311 | 0.404 | 0.076 |
| (1-100) [11-20] | 0.433 | 0.000 | 0.335 | 0.199 | 0.000 | 0.119 | 0.000 | 0.404 | 0.076 |
| (-1010) [-1210] | 0.000 | 0.000 | 0.000 | 0.000 | 0.433 | 0.000 | 0.311 | 0.000 | 0.000 |
| **Pyramidal I <c+a> Slip System** | | | | | | | | | |
| (0-111) [-12-13] | 0.000 | 0.404 | 0.248 | 0.404 | 0.202 | 0.459 | 0.404 | 0.122 | 0.475 |
| (0-11-1) [11-2-3] | 0.202 | 0.404 | 0.000 | 0.202 | 0.404 | 0.306 | 0.202 | 0.122 | 0.356 |
| (01-11) [-12-13] | 0.000 | 0.404 | 0.065 | 0.032 | 0.202 | 0.126 | 0.032 | 0.066 | 0.191 |
| (01-1-1) [-1-12-3] | 0.202 | 0.404 | 0.129 | 0.048 | 0.404 | 0.169 | 0.048 | 0.199 | 0.239 |
| (1-101) [-12-13] | 0.000 | 0.404 | 0.248 | 0.404 | 0.000 | 0.459 | 0.404 | 0.122 | 0.475 |
| (1-101) [-2113] | 0.202 | 0.404 | 0.000 | 0.202 | 0.000 | 0.306 | 0.202 | 0.122 | 0.356 |
| (-1101) [-12-1-3] | 0.000 | 0.404 | 0.065 | 0.032 | 0.000 | 0.126 | 0.032 | 0.066 | 0.191 |
| (-1101) [-211-3] | 0.202 | 0.404 | 0.129 | 0.048 | 0.000 | 0.169 | 0.048 | 0.199 | 0.239 |
| (10-1-1) [-1-12-3] | 0.404 | 0.404 | 0.441 | 0.231 | 0.404 | 0.053 | 0.231 | 0.482 | 0.061 |
| (10-1-1) [-211-3] | 0.404 | 0.404 | 0.441 | 0.231 | 0.202 | 0.053 | 0.231 | 0.482 | 0.061 |
| (-101-1) [11-2-3] | 0.404 | 0.404 | 0.000 | 0.295 | 0.404 | 0.417 | 0.295 | 0.216 | 0.462 |
| (-101-1) [-2113] | 0.404 | 0.404 | 0.000 | 0.295 | 0.202 | 0.417 | 0.295 | 0.216 | 0.462 |